\documentclass[12pt]{article}

\usepackage{jheppub} 
\usepackage{float}
\usepackage{amsmath,bbm,blkarray,bm}
\usepackage[vcentermath]{youngtab}
\usepackage{slashed}
\usepackage{bbold}
\usepackage{booktabs}
\usepackage{multirow}
\usepackage{tikz}
\usepackage{changepage}
\usepackage{makecell}
\usepackage{tikz-feynman}
\usepackage{subcaption}
\usepackage{color}
\usepackage[numbers,sort&compress]{natbib}
\usepackage[T1]{fontenc} 
\usepackage{amsmath}
\usepackage{bbm}

\usepackage{adjustbox}
\usepackage{cancel}

\newcommand{\be} {\begin{equation}}
\newcommand{\ee} {\end{equation}}
\newcommand{\bea} {\begin{eqnarray}}
\newcommand{\eea} {\end{eqnarray}}

\newcommand{\sE}{\mathcal{D}}
\newcommand{\sF}{\mathcal{F}}

\newcommand{\expval}[1]{\left<#1\right>}

\newcommand{\ol}[1]{\overline{#1}}

\allowdisplaybreaks

\title{A Simple Dirac Prescription for Two-Loop Anomalous Dimension Matrices}

\author[a]{Jason~Aebischer,}
\author[a]{Marko Pesut,}
\author[a]{Zachary Polonsky}

\affiliation[a]{Physik-Institut, Universit\"at Z\"urich, CH-8057 Z\"urich, Switzerland}

\emailAdd{jason.aebischer@physik.uzh.ch}
\emailAdd{marko.pesut@physik.uzh.ch}
\emailAdd{zach.polonsky@physik.uzh.ch}

\begin{document}

\abstract{A novel method to treat effects from evanescent operators in next-to-leading order (NLO) computations is introduced. The approach allows, besides further simplifications, to discard  evanescent-to-physical mixing contributions in NLO calculations. The method is independent of the treatments of $\gamma_5$ and can therefore be combined with different renormalization schemes. We illustrate the utility of this result by reproducing literature results of two-loop anomalous dimension matrices for both $|\Delta F| = 1$ and $|\Delta F| = 2$ transitions.}

\maketitle

\section{Introduction}
When performing loop-level computations in perturbative quantum field theories, one frequently encounters
the issue of unphysical ultraviolet (UV) divergences. These UV divergences are handled in a
systematic way using renormalization, where one absorbs the UV divergences into the parameters
of the theory, and reinserts them in physical processes, where the divergences cancel
order-by-order. However, in order to extract the UV divergences, one must first render them finite through the use of
a regularization method, the most commonly used being dimensional regularization where the dimensionality
of spacetime is continued from $d=4$ to $d=4-2\epsilon$, and the limit of $\epsilon\to 0$ is
taken to obtain a physical result. One issue arises when including fermions into the theory, since the
Dirac algebra cannot always be trivially continued to arbitrary dimensionality. In the majority of the literature, this issue is handled by using evanescent operators: operators which
vanish in the physical limit, but account for the fact that operator relations in this limit
may be altered when working in an arbitrary number of dimensions. The standard story of evanescent operators is as follows: In a computation, physical matrix elements may project
onto a set of Dirac structures, $\{\sE\}$. In four dimensions, these structures are not independent
of the set of physical operators, $\{Q\}$, i.e.
\begin{equation}\label{eq:4drelations}
	\sE \, \overset{d=4}{=}\, \sF_4 Q\,,
\end{equation}
where we have specified that this relation is only true in four-dimensional spacetime. When continuing to $d=4-2\epsilon$, the relation in Eq.~\eqref{eq:4drelations} will no longer
hold in general, meaning that the operators $\sE$ and $Q$ are no longer necessarily linearly
dependent. To compensate for this, one adds an additional operator to the basis, known as an \textit{evanescent
operator} (EV) \cite{Buras:1989xd,Dugan:1990df}
\begin{equation}
	E = \sE - \sF Q\,,
\end{equation}
which by definition vanishes from the operator basis when we take the physical limit $\epsilon\to 0$.
Note that the matrix $\sF$ needs not be exactly equal to $\sF_4$ due to the fact that the only requirement
is that $E$ vanishes when $\epsilon\to 0$. Therefore, we may include an arbitrary dependence on positive
powers of $\epsilon$ so that $\sF = \sF_4 + \sigma$ where
\begin{equation}\label{eq:evanH}
	\sigma = \sum_{n = 1}^\infty \epsilon^n \sigma_n\,,
\end{equation}
where the arbitrary constants $\sigma_n$ fix the renormalization scheme \cite{Herrlich:1994kh}. We may therefore write the interaction Hamiltonian in $d$-dimensions as
\begin{equation}
	\mathcal{H}^d_{\text{int}} = C_Q\,Q + C_E\,E\,,
\end{equation}
where $C_Q$ and $C_E$ correspond to the physical and evanescent Wilson coefficients (WCs), respectively. The choice of evanescent operator basis is not unique. In fact, to write down such a basis, one must
first pick a scheme for how to handle Dirac matrix manipulations in $d$-dimensions (hereby referred
to as a ``prescription''), e.g. naive dimensional regularization (NDR), `t Hooft-Veltmann (HV), or
dimensional reduction (DRED). Once a prescription is chosen, it allows one to reduce certain Dirac structures, but not all. The different prescriptions then lead to different irreducible structures, and for each such structure, an evanescent
operator must be introduced.

However, not even the evanescent operators for a given prescription are unique. Since the only requirement
for an evanescent operator is that it vanishes when we take $d\to 4$, we may introduce a general
evanescent-scheme dependence (hereby just called the scheme dependence) like we have done in Eq.~\eqref{eq:evanH}. The key point is the following: any physical observable must be independent of the choice of \textit{both} the
prescription and scheme as alternative choices differ by only $O(\epsilon)$ contributions. However, the set of structures which are 
reducible or irreducible in each prescription are different. This means that, in one prescription, a structure may be
reducible, but in another prescription, the same structure will require an evanescent operator. The issue is that
the treatment of the structure can be substantially different in the two prescriptions: in the second prescription, we must
introduce finite subtractions for this structure to avoid the issue of specifying an infinite number of 
initial conditions for evanescent operators. This finite subtraction does \textit{not} correspond to a choice
of renormalization scheme and cannot be neglected in order to obtain the correct result. In the first prescription, we simply treat the structure as any other and move on. In this article, we examine the notion of different prescriptions and suggest a novel approach for treating structures that can not be trivially reduced to the physical basis. The method is recursive and independent of the chosen renormalization scheme and is hence well suited for automation. 

The rest of the article is structured as follows: In Sec.~\ref{sec:example} an example in scalar QED is considered in order to illustrate different prescriptions and their relations among each other. Furthermore, in Sec.~\ref{sec:pres} a general prescription at the NLO is presented and applied to several examples in the literature. Finally, we conclude in Sec.~\ref{sec:concl}. Details of the two-loop computation from Sec.~\ref{sec:example} are collected in App.~\ref{app:2Lex}.

\section{An Example: Muon decay in scalar QED}\label{sec:example}

As a simple example, we will examine NLO QED corrections to charged lepton decay, mediated by a charged scalar 
particle with interaction Lagrangian
\begin{equation}
	\mathcal{L}\supset -y^L_\ell \big(\bar{\ell} P_L \nu_\ell\big) \phi -
	y^R_\ell \big(\bar{\ell} P_R \nu_\ell\big) \phi +
	\text{h.c.}\,,
\end{equation}
where $P_L$ and $P_R$ denote the left- and right-handed projection operators, respectively. We will only study the left-left decays, as this sector will run independently of the right-right and
mixed sectors under only vector-coupled, parity conserving QED.

We will consider matching this theory onto an effective Hamiltonian with two physical operators which
mix under QED renormalization group running
\begin{equation}
	\mathcal{H}_{\text{EFT}} \supset C_s Q_s + C_t Q_t\,,
\end{equation}
with
\begin{equation}
	Q_s = \big(\bar{\nu}_\mu P_L \mu\big)\big(\bar{e}P_L\nu_e\big), \quad
	Q_t = \big(\bar{\nu}_\mu \sigma^{\mu\nu} P_L \mu\big)\big(\bar{e} \sigma_{\mu\nu}P_L\nu_e\big)\,,
\end{equation}
and $C_s$ and $C_t$ the respective WCs. The purpose of this procedure is to fully compute this process featuring non-trivial 
off-diagonal running in three different ways: first, we will use the standard NDR prescription with scheme-dependent evanescent
operators. Second, we will consider the case where we give no prescription for the treatment of $\gamma_5$ and instead only define evanescent operators
for all Dirac structures which do not immediately reduce with $d$-dimensional Dirac algebra. We will call this prescription the ``no-prescription'' (NP) method. Finally,
we will consider an extension to NDR featuring the so-called ``Greek trick'' \cite{Tracas:1982gp} to define a prescription which covers all
non-trivial Dirac structures and gives no evanescent operators. We note however, that our way of employing the Greek method differs from the conventional one used in the literature. Traditionally, the Greek identities are used in combination with introducing corresponding EVs, as further explained in App.~\ref{app:greek}. To simplify the discussion and comparison with our method we refrain from this approach and simply apply the greek identities, without introducing EVs. This method will be referred to as the ``Greek prescription'' (GP) method. 

Hence, in the following the matching and running computations will be performed in these three prescriptions in order to compare the different results.

\subsection{Matching}

We begin by computing the leading-order (LO) matching onto the effective theory. The tree-level amplitude in the UV theory is
given by 
\begin{equation}
	\mathcal{A}_{\text{UV}}^{(\text{LO})} = \frac{i y_e^L y_\mu^R}{M_S^2}\big(P_L\big)\otimes\big(P_L\big)\,,
\end{equation}
where $M_S$ denotes the mass of the heavy scalar. In the EFT, we expand the WCs as
\begin{equation}
	C_i = C_i^{(0)} + \frac{\alpha}{4\pi}C_i^{(1)} + O\big(\alpha^2\big)\,,
\end{equation}
which gives the LO matching conditions
\begin{equation}
	C_s^{(0)} = -\frac{y_e^Ly_\mu^R}{M_S^2}\,, \quad C_t^{(0)} = 0\,.
\end{equation}
At NLO, the $\overline{\rm MS}$-renormalized amplitude reads in the hard region
\begin{equation}
\label{eq:oneloopUV}
	\begin{split}
		\mathcal{A}_{\text{UV}}^{(\text{NLO})} = -i\frac{\alpha}{4\pi}\frac{y_\mu^Ry_e^L}{M_S^2}
		\Bigg[&\frac{1}{4}\big(P_L\gamma^\mu\gamma^\nu\big)\otimes\big(\gamma_\nu\gamma_\mu P_L\big)
		\Big(\frac{1}{\epsilon} + \frac{3}{2} + \log\frac{\mu^2}{M_S^2}\Big) \\[5pt]
		&- \big(P_L\big)\otimes\big(P_L\big)\Big( 1 + \log\frac{\mu^2}{M_S^2}\Big)\Bigg]\,.
	\end{split}
\end{equation}
The $1/\epsilon$ pole arises from the infrared divergence ($\epsilon < 0$) and matches the corresponding UV pole in the EFT up to a sign, following the method of regions.
Using this method, we can immediately extract the matching conditions from this expression, neglecting the $1/\epsilon$
pole, once we reduce the Dirac algebra to project onto the operators in the EFT.

\subsubsection{NDR Prescription}

In NDR, the structure appearing in the matching can be immediately re-written in terms of the physical operators
\begin{equation}\label{eq:NDRsimplify}
	\big(P_L\gamma^\mu\gamma^\nu\big)\otimes\big(\gamma_\nu\gamma_\mu P_L\big) = 
	(4 - 2\epsilon)\big(P_L\big)\otimes\big(P_L\big) + \big(\sigma^{\mu\nu}P_L\big)\otimes\big(\sigma_{\mu\nu}P_L\big)\,.
\end{equation}
Making this replacement in Eq.~\eqref{eq:oneloopUV} gives the NDR matching conditions
\begin{equation}
	C_s^{(1)}(\mu_0) = 0\,, \quad
	C_t^{(1)}(\mu_0) = \frac{y_\mu^Ry_e^L}{4M_S^2}\Big(\frac{3}{2} + \log\frac{\mu_0^2}{M_S^2}\Big)\,.
\end{equation}

\subsubsection{NP Prescription}

Since in NP, we do not specify any way of treating $\gamma_5$, we require an evanescent operator in order to
reduce the Dirac algebra in the matching. We choose the definition
\begin{equation}
	\ol{E}_1 = \big(\bar{\nu}_\mu P_L \gamma^\mu \gamma^\nu \mu\big)\big(\bar{e}\gamma_\nu \gamma_\mu P_L\nu_e\big)
	- (4 - \epsilon\ol{\sigma}_{s1}) Q_s - (1 - \epsilon \ol{\sigma}_{t1})Q_t\,,
\end{equation}
with arbitrary coefficients $\ol{\sigma}_{s1},\ol{\sigma}_{t1}$.
Note that this implies that the NLO amplitude in the UV theory projects onto the evanescent operator $\ol{E}_1$. 
This projection is irrelevant in the matching since we will end up subtracting any evanescent-to-physical mixing, and therefore we 
need not specify initial conditions for the WCs corresponding to evanescent operators.\footnote{We note, that the evanescent-to-physical mixing contributions are finite. They stem from $\epsilon$-dependent parts resulting from EV insertions, which are multiplying $1/\epsilon$ poles from the corresponding loop diagrams.} Neglecting this projection, we find

\begin{equation}
	\ol{C}_s^{(1)}(\mu_0) = \frac{y_\mu^R y_e^L}{M_S^2}\Big(\frac{1}{2} - \frac{\ol{\sigma}_{s1}}{4}\Big)\,,\quad
	\ol{C}_t^{(1)}(\mu_0) = \frac{y_\mu^R y_e^L}{4M_S^2}\Big(\frac{3}{2} 
		- \ol{\sigma}_{t1} + \log\frac{\mu_0^2}{M_S^2}\Big)\,.
\end{equation}
Clearly, this reproduces the same matching conditions as in NDR if we set $\ol{\sigma}_{s1} = 2$ and $\ol{\sigma}_{t1} = 0$.

\subsubsection{GP Prescription}

In the case of GP, we can again directly evaluate the structure appearing in the matching, giving the same
result as in NDR. Using this, we find the same UV and EFT amplitudes as in NDR and therefore the same matching conditions
\begin{equation}
	\tilde{C}_s^{(1)}(\mu_0) = 0\,, \quad
	\tilde{C}_t^{(1)}(\mu_0) = \frac{y_\mu^Ry_e^L}{4M_S^2}\Big(\frac{3}{2} + \log\frac{\mu_0^2}{M_S^2}\Big)\,.
\end{equation}
Here, one may already expect an inconsistency to arise: in NDR, we will have evanescent operators
which will introduce additional terms corresponding to the finite subtractions for evanescent-to-physical mixing. Therefore,
it seems that the two prescriptions could produce different anomalous dimension matrices (ADMs), yet have the same matching conditions, as discussed in Ref.~\cite{DUGAN1991239}.

\subsection{One-Loop EFT Renormalization}

\begin{figure}[tb]
\centering
\includegraphics[width=3.5cm]{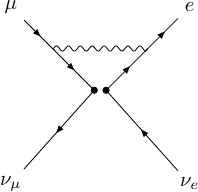}
\caption{One-loop QED vertex corrections to the four-point interaction in the EFT. \label{fig:1L}}
\end{figure}

In order to perform the NLL running from the matching scale to the scale where we evaluate the matrix element, we must perform
the one- and two-loop renormalization in the EFT. We will use $\ol{\text{MS}}$ augmented by finite subtractions for
evanescent-to-physical mixing. The UV poles can be extracted using the methods described in Ref.~\cite{Chetyrkin:1997fm}.
\noindent
At one loop, we only have a single diagram to compute for each operator insertion, which is depicted in Fig.~\ref{fig:1L}. For an operator insertion with general
Dirac structure $(\Gamma_1)\otimes(\Gamma_2)$, we find the result of the diagram
in a general $R_\xi$ gauge
\begin{equation}
\label{eq:master1l}
	\begin{split}
		\hspace{-0.25cm}
		\mathcal{M}^{(1)}_{\text{EFT}} =& -i\frac{\alpha}{4\pi}\,C_{12}\Bigg[
			\frac{1}{4}\big(\Gamma_1\gamma^\mu\gamma^\nu\big)\otimes\big(\gamma_\nu\gamma_\mu\Gamma_2\big)
			+ (\xi - 1)\big(\Gamma_1\big)\otimes\big(\Gamma_2\big)\Bigg]
		\Big(\frac{1}{\epsilon} + \log\frac{\mu^2}{m^2}\Big)\,,
	\end{split}
\end{equation}

\noindent where $m$ is the IR regulator mass (see Ref.~\cite{Chetyrkin:1997fm} for details). The $\xi$-dependence in Eq.~\eqref{eq:master1l} will drop out after the renormalization of the external legs. We have kept the finite
logarithmic term as a check for the cancellation of non-local divergences when inserting operator counterterms. At this point, we must specify how to treat the Dirac algebra in order to project back onto the physical basis and complete the
physical-to-physical and physical-to-evanescent renormalization. In particular, we must know how to treat the two structures
\begin{equation}
	(P_L\gamma^\mu \gamma^\nu)\otimes(\gamma_\nu\gamma_\mu P_L), \quad
	(\sigma^{\mu\nu} P_L\gamma^\alpha\gamma^\beta)\otimes(\gamma_\beta \gamma_\alpha \sigma_{\mu\nu}P_L)
\end{equation}
which arise from the insertions of physical operators. The former structure has already been addressed in the
matching computation. The latter one will now be discussed again in the different prescriptions.

\subsubsection{NDR Prescription}

In NDR, to reduce the second structure, we must introduce an evanescent operator which we choose to take
the form
\begin{equation}
	E_{\text{NDR}} = \big(\bar{\nu}_\mu\gamma^\mu\gamma^\nu\gamma^\alpha\gamma^\beta P_L\mu\big)
		\big(\bar{e}\gamma_\mu\gamma_\nu\gamma_\alpha\gamma_\beta P_L \nu_e\big)
		- (64 - \epsilon \sigma_s) Q_s - (-16 - \epsilon \sigma_t) Q_t\,.
\end{equation}
We then find 
\begin{equation}
	\begin{split}
		\big(\bar{\nu}_\mu\sigma^{\mu\nu} P_L\gamma^\alpha\gamma^\beta\mu\big)\big(\bar{e}\gamma_\beta \gamma_\alpha \sigma_{\mu\nu} P_L\nu_e\big) =& 
		\big(48 - (-16 + \sigma_s)\epsilon\big)Q_s \\[5pt]
		&+ \big(12 - (22 + \sigma_t)\epsilon\big)Q_t + E_{\text{NDR}}\,,
	\end{split}
\end{equation}
Using Eq.~\eqref{eq:master1l}
along with the one-loop lepton field renormalization in the general $R_{\xi}$-gauge
\begin{equation}
	\delta Z_\ell^{(1)} = -\frac{\alpha}{4\pi}\frac{\xi}{\epsilon}\,,
\end{equation}
gives the one-loop renormalization constants
\begin{equation}\label{eq:NDRrenorm1l}
	Z_{QQ}^{(1;1)} = \begin{pmatrix}
		0 & -\frac{1}{4} \\[5pt]
		-12 & -2
	\end{pmatrix}, \qquad
	Z_{QE}^{(1;1)} = \begin{pmatrix}
		0 \\[5pt]
		-\frac{1}{4}
	\end{pmatrix},
\end{equation}
where $Q = \{Q_s, Q_t\}$ and we organize the renormalization constants according to
\begin{equation}
	Z_{ij} = \sum_{n}\sum_m \frac{1}{\epsilon^n}\Big(\frac{\alpha}{4\pi}\Big)^m Z_{ij}^{(m;n)}\,,
\end{equation}
which relates the bare and renormalized operators in the Lagrangian via
\begin{equation}
	\mathcal{O}_i^{(0)} = (\mathbb{1}+Z)_{ij}\mathcal{O}_j\,.
\end{equation}
The operator renormalization constant in eq.~\eqref{eq:NDRrenorm1l} agrees with the findings in \cite{Aebischer:2022aze}.
Only the $\epsilon$-independent piece of the treatment of Dirac algebra enters the one-loop poles, so the
physical-to-physical counterterms will be the same in all prescriptions.

\subsubsection{NP Prescription}

When we do not introduce a prescription for treating $\gamma_5$, we choose a second evanescent operator to treat the
second Dirac structure arising from an insertion of $Q_t$
\begin{equation}
	\ol{E}_2 = \big(\bar{\nu}_\mu \sigma^{\mu\nu}P_L \gamma^\alpha \gamma^\beta \mu\big)
	\big(\bar{e}\gamma_\beta \gamma_\alpha \sigma_{\mu\nu}P_L\nu_e\big)
	-\big(48 - \epsilon\ol{\sigma}_{s2}\big) Q_s - (12 - \epsilon \ol{\sigma}_{t2})Q_t\,.
\end{equation}
This gives, in addition to the physical-to-physical counterterm $Z_{QQ}$ in Eq.~\eqref{eq:NDRrenorm1l}, the physical-to-evanescent
mixing (ordering $\ol{E} = \{\ol{E}_1, \ol{E}_2\}$)
\begin{equation}
	Z_{Q\ol{E}}^{(1;1)} = -\frac{1}{4}\mathbbm{1}_{2\times 2}\,.
\end{equation}

\subsubsection{GP Prescription}

In the case of GP, we can immediately evaluate the two structures appearing in the amplitudes. We find for the
second structure
\begin{equation}
	\big(\sigma^{\mu\nu}P_L\gamma^\alpha\gamma^\beta\big)\otimes\big(\gamma_\beta\gamma_\alpha\sigma_{\mu\nu}P_L\big) = 
	\big(48 - 80\epsilon\big)\big(P_L\big)\otimes\big(P_L\big) 
	+ \big(12 - 14\epsilon\big)\big(\sigma^{\mu\nu}P_L\big)\otimes\big(\sigma_{\mu\nu}P_L\big)\,.
\end{equation}
Since we do not have any evanescent operators in this prescription, we cannot specify any renormalization constant $Z_{Q\tilde{E}}$ corresponding to
physical-to-evanescent mixing.

\subsection{Evanescent-To-Physical Mixing}

In both NDR and NP, we have introduced unphysical evanescent operators whose mixing into the physical sector
must be taken into account. To avoid the need of specifying an infinite number of initial conditions, we must subtract off
the finite evanescent-to-physical mixing to decouple the two sectors. This is done by inserting evanescent operators into
Eq.~\eqref{eq:master1l} and keeping the finite pieces.

\subsubsection{NDR Prescription}

In the case of NDR, we must introduce one additional evanescent operator which appears when inserting the four-gamma
structure into Eq.~\eqref{eq:master1l}. We choose this operator to be
\begin{equation}
	\begin{split}
		E'_{\text{NDR}} =& \big(\bar{\nu}_\mu \gamma^\mu \gamma^\nu \gamma^\alpha \gamma^\beta \gamma^\sigma \gamma^\rho P_L\mu\big)
		\big(\bar{e} \gamma_\mu \gamma_\nu \gamma_\alpha \gamma_\beta \gamma_\sigma \gamma_\rho P_L\nu_e\big)\\[5pt]
		&\quad- (1024 - \epsilon \sigma'_s)Q_s - (-256 - \epsilon \sigma'_t)Q_t\,.
	\end{split}
\end{equation}
With this definition we find the relation:
\begin{equation}
	\begin{split}
		&\big(\bar{\nu}_\mu \gamma^\mu \gamma^\nu \gamma^\alpha \gamma^\beta \gamma^\sigma \gamma^\rho P_L\mu\big)
		\big(\bar{e}\gamma_\rho\gamma_\sigma  \gamma_\mu \gamma_\nu \gamma_\alpha \gamma_\beta  P_L\nu_e\big) = \\
	&(-512 + (- 1024 - 8\sigma_s + \sigma_s')\epsilon)Q_s
		+(-128 +(512 - 8\sigma_t + \sigma_t')\epsilon)Q_t \\[5pt]
		&+ (8 - 36\epsilon)E_{\text{NDR}} - E'_{\text{NDR}}\,.
\end{split}
\end{equation}
Also,
\begin{equation}
	\begin{split}
		&\big(\bar{\nu}_\mu \sigma^{\mu \nu} \gamma^\alpha \gamma^\beta \gamma^\sigma \gamma^\rho P_L\mu\big)
		\big(\bar{e}\gamma_\alpha \gamma_\beta  \gamma_\sigma \gamma_\rho \sigma_{\mu \nu} P_L\nu_e\big) = \\
	&(-768+(-896-4\sigma_s+\sigma_s^\prime)\epsilon)Q_s+(-64+(480-4\sigma_t+\sigma_t^\prime)\epsilon)Q_t+(4-34\epsilon)E_{\text{NDR}}-E'_{\text{NDR}}\,.
\end{split}
\end{equation}
After inserting $E_{\text{NDR}}$ into the one-loop diagram and projecting onto the physical basis, we find the finite counterterm
\begin{equation}
	Z_{EQ}^{(1;0)} = \begin{pmatrix}
		160 + 5\sigma_s - 12\sigma_t -\frac{\sigma_s'}{4} &&
		-40 - \frac{\sigma_s}{4} + 3\sigma_t - \frac{\sigma_t'}{4}
	\end{pmatrix}\,.
\end{equation}
Note that we do not need to insert $E'_{\text{NDR}}$ since the physical operators do not mix into
this evanescent operator at the considered order.

\subsubsection{NP Prescription}

Upon inserting our evanescent operators in NP, we will require two additional evanescent operators, which we
choose to be
\begin{equation}\label{eq:twoloopNPEs}
	\begin{split}
		\ol{E}'_1 =& \big(\bar{\nu}_\mu P_L\gamma^\mu \gamma^\nu \gamma^\alpha \gamma^\beta \mu\big)
			\big(\bar{e}\gamma_\beta \gamma_\alpha \gamma_\nu \gamma_\mu P_L\nu_e\big)- (64 - \epsilon \ol{\sigma}'_{s1})Q_s - (16 - \epsilon \ol{\sigma}'_{t1})Q_t\,,\\[5pt]
		\ol{E}'_2 =& \big(\bar{\nu}_\mu \sigma^{\mu\nu} P_L\gamma^\alpha \gamma^\beta \gamma^\sigma \gamma^\rho \mu\big)
			\big(\bar{e}\gamma_\rho \gamma_\sigma \gamma_\beta \gamma_\alpha \sigma_{\mu\nu} P_L\nu_e\big) - \big(768 - \epsilon \ol{\sigma}'_{s2}\big) Q_s - \big(192 - \epsilon \ol{\sigma}'_{t2}\big)Q_t\,.
	\end{split}
\end{equation}
With these, we can insert $\ol{E}_1$ and $\ol{E}_2$ into the one-loop diagram to find the finite subtractions

\begin{equation}
	\hspace{-0.5cm}
	Z_{\ol{E} Q}^{(1;0)} = 
		\begin{pmatrix}
		 - 2 \ol{\sigma}_{s1} + \frac{\ol{\sigma}'_{s1}}{4} 
			- \frac{\ol{\sigma}_{s2}}{4} -12\ol{\sigma}_{t1} &&
		- \frac{\ol{\sigma}_{s1}}{4} - 4\ol{\sigma}_{t1} + \frac{\ol{\sigma}'_{t1}}{4} - \frac{\ol{\sigma}_{t2}}{4} \\[5pt]
		 - 12 \ol{\sigma}_{s1} -4\ol{\sigma}_{s2} + \frac{\ol{\sigma}'_{s2}}{4} -12\ol{\sigma}_{t2} && 
		- \frac{\ol{\sigma}_{s2}}{4} - 12\ol{\sigma}_{t1} -6\ol{\sigma}_{t2} + \frac{\ol{\sigma}'_{t2}}{4}
	\end{pmatrix}\,.
\end{equation}

Here, we point out the (somewhat obvious) observation that these counterterms are purely scheme-dependent. This is due to the fact that the scheme-independent pieces, i.e. the purely four-dimensional parts, cancel trivially due to the requirement that the structures appearing in loop diagrams must reduce to the four-dimensional counterparts when $d\to 4$. We are therefore only left with pieces where the
$1/\epsilon$ poles of the loop integrals hit the $\epsilon$ terms in the definition of the evanescent operators. This pure scheme dependence is much more clear
in NP than in NDR due to the fact that, in NDR, the Dirac matrices appearing in the amplitudes are re-ordered to project onto the chosen evanescent operators, thereby introducing additional
$O(\epsilon)$ terms from $d$-dimensional Dirac algebra. In NP, on the other hand, no such re-arranging occurs, making the pure scheme-dependence more explicit.

\subsection{Anomalous Dimension Matrix}

The next step is the computation of the one- and two-loop anomalous dimension matrix corresponding to physical-to-physical mixing.
The one-loop ADM can immediately be found from the one-loop counterterms. Expanding the ADM as
\begin{equation}
	\gamma = \frac{\alpha}{4\pi}\gamma^{(0)} + \Big(\frac{\alpha}{4\pi}\Big)^2 \gamma^{(1)} + O(\alpha^3)\,,
\end{equation}
we find
\begin{equation}
	\gamma^{(0)} = \begin{pmatrix}
		0 && -\frac{1}{2} \\[5pt]
		-24 && -4
	\end{pmatrix}\,,
\end{equation}
in all prescriptions.

The two-loop ADM is found in the $\ol{\text{MS}}$-scheme by computing only the UV-divergent pieces of all
two-loop diagrams with physical operator insertions. The UV poles can be extracted using the methods presented in
Ref.~\cite{Chetyrkin:1997fm}. We must additionally compute all one-loop diagrams with single counterterm insertions
in order to subtract sub-divergences in the two-loop diagrams. In total, this amounts to the computation of thirteen diagrams,
up to Dirac algebra: seven true two-loop diagrams, five dimension-four operator counterterm insertions, and one dimension-six
operator counterterm insertion. The necessary counterterms and the pole structure of the corresponding diagrams are collected in App.~\ref{app:2Lex}.

\subsubsection{NDR Prescription}

In NDR, using the previously-defined evanescent operators, we find the (local and gauge-invariant) counterterms
\begin{equation}
	\begin{split}
		&Z_{QQ,\text{NDR}}^{(2;2)} = \begin{pmatrix}
			\frac{3}{2} && -\frac{1}{12} \\[5pt]
			-4 && \frac{5}{6}
		\end{pmatrix}, \\[10pt]
		&Z_{QQ,\text{NDR}}^{(2;1)} = \begin{pmatrix}
			-\frac{21}{4} -\frac{\sigma_s}{32} &&& 
				-\frac{23}{72} - \frac{\sigma_t}{32} \\[5pt]
			\frac{70}{3} -\frac{13\sigma_s}{24} + 3\sigma_t + \frac{\sigma_s'}{32} &&&
				\frac{283}{36} + \frac{\sigma_s}{16} - \frac{\sigma_t}{24} + \frac{\sigma_t'}{32}
		\end{pmatrix},
	\end{split}
\end{equation}
which gives the two-loop anomalous dimension matrix
\begin{equation}
	\gamma^{(1)}_{\text{NDR}} = \begin{pmatrix}
		-21 - \frac{\sigma_s}{8} &&& -\frac{23}{18} - \frac{\sigma_t}{8} \\[5pt]
		\frac{520}{3} + \frac{\sigma_s}{3} + 6\sigma_t &&& 
			\frac{103}{9} + \frac{\sigma_s}{8} + \frac{4\sigma_t}{3}
	\end{pmatrix}.
\end{equation}
As additional checks, we explicitly verified the relation between the $1/\epsilon^2$ poles of $Z_{QQ}^{(2;2)}$ to
the $1/\epsilon$ poles of $Z_{QQ}^{(1;1)}$ and we also find that the ADM is independent of $\sigma_s'$ and $\sigma_t'$,
as must be the case in order for the scheme-dependence to properly cancel with the one-loop matching and matrix elements.

\subsubsection{NP Prescription}

Performing the same computation using NP, we find the counterterms

\begin{equation}
	\begin{split}
		&\ol{Z}_{QQ}^{(2;2)} = \begin{pmatrix}
			\frac{3}{2} && -\frac{1}{12} \\[5pt]
			-4 && \frac{5}{6}
		\end{pmatrix}, \\[10pt]
		&\ol{Z}_{QQ}^{(2;1)} = \begin{pmatrix}
			-\frac{77}{12} + \frac{7}{12}\ol{\sigma}_{s1} 
				+ 3\ol{\sigma}_{t1} - \frac{\ol{\sigma}_{s1}'}{32}&&& 
			\frac{11}{36} + \frac{\ol{\sigma}_{s1}}{16} + \frac{13\ol{\sigma}_{t1}}{12} - \frac{\ol{\sigma}_{t1}'}{32} \\[5pt]
			\frac{44}{3} + \frac{7\ol{\sigma}_{s2}}{12} + 3\ol{\sigma}_{t2} - \frac{\ol{\sigma}_{s2}'}{32} &&&
			- \frac{143}{36} + \frac{\ol{\sigma}_{s2}}{16} + \frac{13\ol{\sigma}_{t2}}{12} - \frac{\ol{\sigma}_{t2}'}{32}
		\end{pmatrix},
	\end{split}
\end{equation}
giving the two-loop ADM
\begin{equation}
	\gamma_{\text{NP}}^{(1)} = \begin{pmatrix}
		-\frac{77}{3} + \frac{4}{3}\ol{\sigma}_{s1} 
			+ 6\ol{\sigma}_{t1} - \frac{\ol{\sigma}_{s2}}{8}  &&& 
		\frac{11}{9} + \frac{\ol{\sigma}_{s1}}{8} + \frac{7\ol{\sigma}_{t1}}{3} - \frac{\ol{\sigma}_{t2}}{8} \\[5pt]
		\frac{176}{3} - 6\ol{\sigma}_{s1} + \frac{\ol{\sigma}_{s2}}{3} + 6\ol{\sigma}_{t2} &&& 
			-\frac{143}{9} - 6\ol{\sigma}_{t1} + \frac{\ol{\sigma}_{s2}}{8} + \frac{4\ol{\sigma}_{t2}}{3}
	\end{pmatrix}\,.
\end{equation}
We again see the independence of the two-loop evanescent scheme, indicating that scheme-dependence will
properly cancel with the one-loop matching and matrix elements.

\subsubsection{GP Prescription}

In GP, we require two additional relations in order to reduce the Dirac algebra in $d$-dimensions. We find
\begin{equation}
	\begin{split}
		&(P_L\gamma^\mu\gamma^\nu\gamma^\alpha\gamma^\beta)\otimes(\gamma_\beta\gamma_\alpha\gamma_\nu\gamma_\mu P_L)= \\[5pt]
			&\hspace{3cm}(64 - 96\epsilon)(P_L)\otimes(P_L) 
			+ (16 - 16\epsilon)(\sigma^{\mu\nu}P_L)\otimes(\sigma_{\mu\nu}P_L)\,, \\[5pt]
		&(\sigma^{\mu\nu}P_L\gamma^\alpha\gamma^\beta\gamma^\sigma\gamma^\rho)
			\otimes(\gamma_\rho\gamma_\sigma\gamma_\beta\gamma_\alpha\sigma_{\mu\nu}P_L) = \\[5pt]
			&\hspace{3cm}(768 - 2048\epsilon)(P_L)\otimes(P_L) 
			+ (192 - 416\epsilon)(\sigma^{\mu\nu}P_L)\otimes(\sigma_{\mu\nu}P_L)\,.
	\end{split}
\end{equation}
These relations produce the counterterms
\begin{equation}
	\tilde{Z}_{QQ}^{(2;2)} = \begin{pmatrix}
			\frac{3}{2} && -\frac{1}{12} \\[5pt]
			-4 && \frac{5}{6}
		\end{pmatrix}, \quad 
	\tilde{Z}_{QQ}^{(2;1)} = \begin{pmatrix}
			-\frac{33}{4} && -\frac{5}{72} \\[5pt]
			\frac{118}{3} && \frac{115}{36}
	\end{pmatrix},
\end{equation}
giving the ADM
\begin{equation}
	\gamma^{(1)}_{\text{GP}} = \begin{pmatrix}
		-33 && -\frac{5}{18} \\[5pt]
		\frac{472}{3} && \frac{115}{9}
	\end{pmatrix}.
\end{equation}
Here, we note that $\gamma^{(1)}_{\text{GP}}$ exactly coincides with $\gamma^{(1)}_{\text{NDR}}$ when setting
$\sigma_s = 96$, $\sigma_t = -8$, $\sigma_s' = 2944$, and $\sigma_t' = -352$ and $\gamma^{(1)}_{\text{NP}}$ when 
setting $\ol{\sigma}_{s1} = 2$, $\ol{\sigma}_{t1} = 0$, $\ol{\sigma}_{s2} = 80$, $\ol{\sigma}_{t2} = 14$, $\ol{\sigma}_{s1}' = 96$, 
$\ol{\sigma}_{t1}' = 16$, $\ol{\sigma}_{s2}' = 2048$, and $\ol{\sigma}_{t2}' = 416$, all consistent with the 
relations found using the Greek prescription. We also remark that, for this choice of NDR and NP scheme constants, $Z_{EQ}$ vanishes
for both prescriptions. We defer to Sec.~\ref{sec:discuss} for further discussion of the significance of this observation.

\subsection{Discussion of the Example}\label{sec:discuss}

In this example, we have performed the computation of heavy-scalar-mediated muon decay to NLO in QED using three
different prescriptions for the treatment of Dirac algebra in $d$-dimensions. Our goal was to gain further insight
into the relevance of evanescent operators in such computations. All three prescriptions not only generated different
forms of the evanescent operators, but also different numbers of evanescent operators as well as free parameters
used to fix the scheme. Despite this, all three prescriptions produce the same result when a consistent evanescent scheme is chosen.

This is particularly surprising when considering the Greek prescription, which did not introduce \textit{any} evanescent operators.
Upon closer inspection, it is clear why the GP reproduces the same result as the other two prescriptions: for the particular
choices of $\sigma_i^{(\prime)}$ and $\ol{\sigma}_{i}^{(\prime)}$ which give the same Dirac algebra replacements as those found
using the GP, the evanescent-to-physical counterterms vanish. So, in GP, the problematic mixing of evanescent structures
into the physical sector is automatically handled simply by the replacement we use for $d$-dimensional Dirac algebra.
It seems that one can circumvent the arguments of Ref.~\cite{DUGAN1991239} and ignore the effects of evanescent operators
in the calculation of ADMs if one can ensure \textit{a priori} the vanishing of evanescent-to-physical mixing. 

This becomes even clearer by a simple re-interpretation of evanescent operators in general. Consider a set of operators,
$\sE_i$ which reduce to the set of physical operators, $Q_i$ in four dimensions via
\begin{equation}\label{eq:4drel}
	\sE_i - \big(\sF_4\big)_{ij}\,Q_j \overset{d=4}{=} 0\,.
\end{equation}
When upgrading to $d$-dimensions, we will assert that the following relation, which reduces to the four-dimensional one for
$d\to 4$, holds
\begin{equation}\label{eq:wrongdrel}
	\sE_i - \sF_{ij}\,Q_j = 0\,,
\end{equation}
where
\begin{equation}
	\sF_{ij} = \big(\sF_4\big)_{ij} + \sum_n\epsilon^n\,\sigma^{(n)}_{ij}\,.
\end{equation}
We next consider inserting Eq.~\eqref{eq:wrongdrel} into a one-loop amplitude and we find
\begin{equation}\label{eq:shiftedrel}
	\expval{\sE_i}^{(1)} - \sF_{ij}\expval{Q_j}^{(1)} = \Delta_{ij}\expval{Q_j}^{(0)}\,,
\end{equation}
where $\Delta_{ij} \neq 0$ and moreover is finite. The latter point is very important since we can immediately take the limit
of Eq.~\eqref{eq:shiftedrel} as $d\to 4$, and we find a result which clearly violates Eq.~\eqref{eq:4drel}.

We see that, when using a general prescription with dimensional regularization, loop-effects explicitly spoil four-dimensional
Dirac algebra, reminiscent of loop effects inducing gauge anomalies. However, unlike gauge anomalies, the violation of four-dimensional Dirac
relations is entirely local, arising only from UV poles. We can therefore erase these problematic shifts by introducing
additional, local operators into our basis and defining a subtraction scheme to exactly cancel these terms. These are
precisely the evanescent operators.

\section{A Prescription free of Evanescent-to-Physical mixing}\label{sec:pres}
In this section, we introduce a general methodology for choosing evanescent operator definitions to guarantee the cancellation of
evanescent mixing into the physical sector for four-fermion operators. We then apply this technique to two non-trivial examples in the
Weak Effective Theory (WET) and reproduce known results from the literature.

\subsection{General discussion}

As exemplified by the results of Sec.~\ref{sec:example}, it is clear that, if we can ensure that no evanescent-to-physical
mixing arises when inserting evanescent operators into loop diagrams, then the effect of evanescent operators on physical ADMs is equivalent
to simple ``replacement rules'' like those used in the Greek prescription. More importantly, loop diagrams with evanescent operator insertions
need not be computed, as no additional finite subtractions are necessary, thereby sequestering the evanescent sector to only mix with itself.\\
To begin, we consider a set of physical four-fermion operators defined as
\begin{equation}\label{eq:physOp}
	Q_{ij;a} = (\bar{\psi}_{a_1}\Gamma^a_i\psi_{a_2})(\bar{\psi}_{a_3}\Gamma^a_j\psi_{a_4})\,,
\end{equation}
where $a_i$ are flavor/color indices. Inserting Eq.~\eqref{eq:physOp} into a one-loop diagram results in
\begin{equation}\label{eq:loopStrucPhysIn}
	\expval{Q_{ij;a}}^{(1)} \propto (\bar\psi_{b_1}\Gamma_{\alpha_1}\Gamma^a_i\Gamma_{\alpha_2}\psi_{b_2})
		(\bar\psi_{b_3}\Gamma_{\alpha_3}\Gamma^a_j\Gamma_{\alpha_4}\psi_{b_4})\,,
\end{equation}
where we accounted for the fact that the one-loop insertions can change the color and flavor structure of the external legs.
In $d$-dimensions, this structure does not reduce to any appearing in the physical basis, and we must therefore introduce an evanescent operator.
As previously discussed, the choice of evanescent operators is not unique, so we make the choice
\begin{equation}\label{eq:evanOp1l}
	E^{(1);a\to\,b}_{\alpha_1i\alpha_2;\alpha_3j\alpha_4} = 
		(\bar{\psi}_{b_1}\Gamma_{\alpha_1}\Gamma^a_i\Gamma_{\alpha_2}\psi_{b_2})(\bar{\psi}_{b_3}\Gamma_{\alpha_3}\Gamma^a_j\Gamma_{\alpha_4}\psi_{b_4})
		- C^{k\ell;c}_{\alpha_1i\alpha_2;\alpha_3j\alpha_4;a\to b}\,Q_{k\ell;c}\,,
\end{equation}
where $C^{k\ell;c}_{\alpha_1i\alpha_2;\alpha_3j\alpha_4;a\to b}$ contains not only the coefficients corresponding to the reduction of Eq.~\eqref{eq:loopStrucPhysIn}
in four dimensions, but also arbitrary $O(\epsilon^n)$ terms for $n\ge1$. The super/subscript $a\to b$ refers to the fact that Eq.~\eqref{eq:evanOp1l} arises
from inserting operators with physical flavor/color structure $\{a_\ell\}$ and projecting onto structure $\{b_\ell\}$.\\
With this, we now must account for the fact that the evanescent operators introduced in Eq.~\eqref{eq:evanOp1l} can mix into the physical sector, thereby requiring finite subtractions. Inserting Eq.~\eqref{eq:evanOp1l} into one-loop diagrams will give the new Dirac structure
\begin{equation}\label{eq:loopStrucEvIn}
	\begin{split}
		\expval{E^{(1);a\to b}_{\alpha_1i\alpha_2;\alpha_3j\alpha_4}}^{(1)}\propto&
		(\bar\psi_{c_1}\Gamma_{\beta_1}\Gamma_{\alpha_1}\Gamma^a_i\Gamma_{\alpha_2}\Gamma_{\beta_2}\psi_{c_2})
		(\bar\psi_{c_3}\Gamma_{\beta_3}\Gamma_{\alpha_3}\Gamma^a_j\Gamma_{\alpha_4}\Gamma_{\beta_4}\psi_{c_4})\\[0.5em]
		&- C^{k\ell;d}_{\alpha_1i\alpha_2;\alpha_3j\alpha_4;a\to b}
		(\bar{\psi}_{c_1}\Gamma_{\beta_1}\Gamma^d_k\Gamma_{\beta_2}\psi_{c_2})(\bar{\psi}_{c_3}\Gamma_{\beta_3}\Gamma^d_\ell\Gamma_{\beta_4}\psi_{c_4})\,.
	\end{split}
\end{equation}
Note that, since we have explicitly separated the Dirac structures arising from the loop insertions, i.e. $\Gamma_{\beta_i}$, from the
operator Dirac structures ($\Gamma_{\alpha_k}\Gamma_i\Gamma_{\alpha_\ell}$ and $\Gamma_i$, respectively for the two terms), the two terms in
Eq.~\eqref{eq:loopStrucEvIn} receive the same proportionality constant from e.g. loop integration and reduction of $\Gamma_{\beta_i}$ in the
chosen scheme for the mass dimension-four sector.\\
For the first term in Eq.~\eqref{eq:loopStrucEvIn}, we must again introduce an additional evanescent operator to treat the irreducible Dirac structure. We then choose
\begin{equation}\label{eq:evaOp2l}
	\begin{split}
		E^{(2);a\to c}_{\beta_1\alpha_1i\alpha_2\beta_2;\beta_3\alpha_3j\alpha_4\beta_4} =& 
		(\bar\psi_{c_1}\Gamma_{\beta_1}\Gamma_{\alpha_1}\Gamma^a_i\Gamma_{\alpha_2}\Gamma_{\beta_2}\psi_{c_2})
		(\bar\psi_{c_3}\Gamma_{\beta_3}\Gamma_{\alpha_3}\Gamma^a_j\Gamma_{\alpha_4}\Gamma_{\beta_4}\psi_{c_4})\\[0.5em]
		&- K^{k\ell;d}_{\beta_1\alpha_1i\alpha_2\beta_2;\beta_3\alpha_3j\alpha_4\beta_4; a\to c}\,Q_{k\ell;d}\,.
	\end{split}
\end{equation}
However, the second term in Eq.~\eqref{eq:loopStrucEvIn} is just a one-loop physical operator insertion and can then be re-written in terms of Eq.~\eqref{eq:evanOp1l},
\begin{equation}
	(\bar{\psi}_{c_1}\Gamma_{\beta_1}\Gamma^d_k\Gamma_{\beta_2}\psi_{c_2})(\bar{\psi}_{c_3}\Gamma_{\beta_3}\Gamma^d_\ell\Gamma_{\beta_4}\psi_{c_4}) =
	E^{(1); d\to c}_{\beta_1k\beta_2;\beta_3\ell\beta_4} + C^{mn;f}_{\beta_1k\beta_2;\beta_3\ell\beta_4;d\to c}Q_{mn;f}\,.
\end{equation}
With this, the one-loop evanescent insertion is given by
\begin{equation}\label{eq:reductionCondition}
	\begin{split}
		\expval{E^{(1);a\to b}_{\alpha_1i\alpha_2;\alpha_3j\alpha_4}}^{(1)}&\propto
		E^{(2);a\to c}_{\beta_1\alpha_1i\alpha_2\beta_2;\beta_3\alpha_3j\alpha_4\beta_4} 
		- C^{k\ell;d}_{\alpha_1i\alpha_2;\alpha_3j\alpha_4;a\to b}E^{(1);d\to c}_{\beta_1k\beta_2;\beta_3\ell\beta_4}\\[0.5em]
		&+ \big(K^{mn;f}_{\beta_1\alpha_1 i\alpha_2\beta_2;\beta_3\alpha_3j\alpha_4\beta_4; a\to c}
			- C^{k\ell;d}_{\alpha_1i\alpha_2;\alpha_3j\alpha_4;a\to b}C^{mn;f}_{\beta_1k\beta_2;\beta_3\ell\beta_4;d\to c}\big)Q_{mn;f}\,.
	\end{split}
\end{equation}
It then becomes clear that the evanescent-to-physical mixing vanishes if we fix the scheme of Eq.~\eqref{eq:evaOp2l} such that
\begin{equation}\label{eq:nonFierz}
	K^{mn;f}_{\beta_1\alpha_1 i\alpha_2\beta_2;\beta_3\alpha_3j\alpha_4\beta_4; a\to c}
			= C^{k\ell;d}_{\alpha_1i\alpha_2;\alpha_3j\alpha_4;a\to b}C^{mn;f}_{\beta_1k\beta_2;\beta_3\ell\beta_4;d\to c}\,.
\end{equation}
This procedure simplifies in the case that the reduction of structures onto the physical basis in four-dimensions does not require a rearrangement 
of the external fields, i.e.
\begin{equation}
	C^{k\ell;d}_{\alpha_1i\alpha_2;\alpha_3j\alpha_4;a\to b} = C^{k\ell}_{\alpha_1i\alpha_2;\alpha_3j\alpha_4;a\to b}\delta_{bd}\,.
\end{equation}
In this case (which we will refer to as the non-Fierz case), the four-dimensional reduction only changes the Dirac structure and is completely independent
of the particular choice of external legs. We can then define universal evanescent tensor products without reference to external states
\begin{equation}
	\begin{split}
		&E^{(1)}_{\alpha_1i\alpha_2;\alpha_3j\alpha_4} = (\Gamma_{\alpha_1}\Gamma_i\Gamma_{\alpha_2})\otimes(\Gamma_{\alpha_3}\Gamma_j\Gamma_{\alpha_4})
			- C^{k\ell}_{\alpha_1i\alpha_2;\alpha_3j\alpha_4}(\Gamma_k)\otimes(\Gamma_\ell)\,,\\[0.5em]
		&E^{(2)}_{\beta_1\alpha_1i\alpha_2\beta_2;\beta_3\alpha_3j\alpha_4\beta_4} = 
			(\Gamma_{\beta_1}\Gamma_{\alpha_1}\Gamma_i\Gamma_{\alpha_2}\Gamma_{\beta_2})\otimes(\Gamma_{\beta_3}\Gamma_{\alpha_3}\Gamma_j\Gamma_{\alpha_4}\Gamma_{\beta_4})
			\\[0.5em]
			&\hspace{4cm}- K^{k\ell}_{\beta_1\alpha_1i\alpha_2\beta_2;\beta_3\alpha_3j\alpha_4\beta_4}(\Gamma_k)\otimes(\Gamma_\ell)\,.
	\end{split}
\end{equation}
Similar to before, we can require that the evanescent mixing into the physical sector vanishes, giving the requirement
\begin{equation}
	K^{mn}_{\beta_1\alpha_1i\alpha_2\beta_2;\beta_3\alpha_3j\alpha_4\beta_4} 
		= C^{k\ell}_{\alpha_1i\alpha_2;\alpha_3j\alpha_4}C^{mn}_{\beta_1k\beta_2;\beta_3\ell\beta_4}\,.
\end{equation}
As we have already discussed, once we have guaranteed that this mixing vanishes, the effect of the evanescent operators on physical ADMs is equivalent to that of
replacement rules taking
\begin{equation}\label{eq:reductions}
	\begin{split}
		&(\Gamma_{\alpha_1}\Gamma_i\Gamma_{\alpha_2})\otimes(\Gamma_{\alpha_3}\Gamma_j\Gamma_{\alpha_4})\to
		C^{k\ell}_{\alpha_1i\alpha_2;\alpha_3j\alpha_4}(\Gamma_k)\otimes(\Gamma_\ell)\,, \\[0.5em]
		&(\Gamma_{\beta_1}\Gamma_{\alpha_1}\Gamma_i\Gamma_{\alpha_2}\Gamma_{\beta_2})\otimes(\Gamma_{\beta_3}\Gamma_{\alpha_3}\Gamma_j\Gamma_{\alpha_4}\Gamma_{\beta_4})
		\to C^{k\ell}_{\alpha_1i\alpha_2;\alpha_3j\alpha_4}C^{mn}_{\beta_1k\beta_2;\beta_3\ell\beta_4}(\Gamma_m)\otimes(\Gamma_n)\,.
	\end{split}
\end{equation}
However, the second line of Eq.~\eqref{eq:reductions} simply corresponds to the recursive application of the first line, as is required for such a
prescription to be self-consistent.

In our example of scalar-mediated muon decay, choosing this particular scheme corresponds to fixing the two relations generated by
one-loop physical operator insertions
\begin{equation}
	\begin{split}
		(P_L\gamma^\mu\gamma^\nu)\otimes(\gamma_\nu\gamma_\mu P_L) \to& 
			(4 - \epsilon \ol{\sigma}_{s1})(P_L)\otimes(P_L) + 
			(1 - \epsilon\ol{\sigma}_{t1})(\sigma^{\mu\nu}P_L)\otimes(\sigma_{\mu\nu}P_L)\,, \\[5pt]
		(\sigma^{\mu\nu}P_L\gamma^\alpha\gamma^\beta)\otimes(\gamma_\beta\gamma_\alpha\sigma_{\mu\nu}P_L)
		\to& (48 - \epsilon\ol{\sigma}_{s2})(P_L)\otimes(P_L) +
		(12 - \epsilon\ol{\sigma}_{t2})(\sigma^{\mu\nu}P_L)\otimes(\sigma_{\mu\nu}P_L)\,.
	\end{split}
\end{equation}
These relations, along with the Lorentz- and gauge-invariant condition $\{\gamma^\mu,\gamma^\nu\} = 2\eta^{\mu\nu}$,
completely fixes the Dirac reductions for the problem at hand. For example, using these relations, we find 
\begin{equation}
	\begin{split}
		(P_L\gamma^\mu\gamma^\nu\gamma^\alpha\gamma^\beta)\otimes(\gamma_\beta\gamma_\alpha\gamma_\nu\gamma_\mu P_L)
		\to& \big(64 - \{8\ol{\sigma}_{s1} + \ol{\sigma}_{s2} + 48\ol{\sigma}_{t1}\}\epsilon\big)(P_L)\otimes(P_L) \\[5pt]
		&\hspace{-2cm}+ \big(16 - \{\ol{\sigma}_{s1} + 16\ol{\sigma}_{t1} + \ol{\sigma}_{t2}\}\epsilon\big)(\sigma^{\mu\nu}P_L)\otimes(\sigma_{\mu\nu}P_L)
		\,, \\[5pt]
		(\sigma^{\mu\nu}P_L\gamma^\alpha\gamma^\beta\gamma^\sigma\gamma^\rho)
			\otimes(\gamma_\rho\gamma_\sigma\gamma_\beta\gamma_\alpha\sigma_{\mu\nu}P_L) \to&
		\big(768 - \{48\ol{\sigma}_{s1} + 16 \ol{\sigma}_{s2} + 48\ol{\sigma}_{t2}\}\epsilon\big)(P_L)\otimes(P_L) \\[5pt]
		&\hspace{-2cm}+ \big(192 - \{\ol{\sigma}_{s2} + 48\ol{\sigma}_{t1} + 24\ol{\sigma}_{t2}\}\epsilon)
		(\sigma^{\mu\nu}P_L)\otimes(\sigma_{\mu\nu}P_L)\,.
	\end{split}
\end{equation}
Notice that this is equivalent to fixing the $\ol{\sigma}_i'$ in Eq.~\eqref{eq:twoloopNPEs} to the values in the curly brackets. As expected,
the evanescent-to-physical mixing exactly vanishes with this choice of scheme.

It is worth emphasizing that one must be extremely cautious when using the NDR prescription due to the fact that pushing factors of $\gamma_5$ across
the fermion line, or mixing the Dirac structures from the mass-dimension four and operator insertions changes the Dirac structures in each fermion line, 
thus obscuring the necessary cancellation in Eq.~\eqref{eq:reductionCondition}. For example, if we fix the NDR relation
\begin{equation}\label{eq:NDRreduce}
	\begin{split}
		(\gamma^\mu\gamma^\nu\gamma^\alpha\gamma^\beta P_L)\otimes(\gamma_\mu\gamma_\nu\gamma_\alpha\gamma_\beta P_L) \to&
		(64 - \epsilon\sigma_s)(P_L)\otimes(P_L) \\[5pt]
		&+ (-16 - \epsilon\sigma_t)(\sigma^{\mu\nu}P_L)\otimes(\sigma_{\mu\nu}P_L)\,,
	\end{split}
\end{equation}
then the one-loop insertions of the left-hand side of this relation will generate structures like
\begin{equation}
	(\gamma^\mu\gamma^\nu\gamma^\alpha\gamma^\beta P_L \gamma^\sigma\gamma^\rho)\otimes
	(\gamma_\rho\gamma_\sigma\gamma_\mu\gamma_\nu\gamma_\alpha\gamma_\beta P_L)\,.
\end{equation}
If we immediately use Eq.~\eqref{eq:NDRreduce}, then we will see the explicit cancellation of the shifts to the
four-dimensional Dirac algebra, but if we first use the naive anticommutation of $\gamma_5$ and re-order the Dirac matrices
in both lines, we will now obtain structures like
\begin{equation}\label{eq:wrongreduce}
	(\gamma^\mu\gamma^\nu\gamma^\alpha\gamma^\beta\gamma^\sigma\gamma^\rho P_L)\otimes
	(\gamma_\mu\gamma_\nu\gamma_\alpha\gamma_\beta\gamma_\sigma\gamma_\rho P_L)\,,
\end{equation}
However, this spoils the overall proportionality seen in Eq.~\eqref{eq:reductionCondition} since the re-ordering will be different between the two terms, thus violating four-dimensional Dirac algebra relations and requiring finite subtractions.
Indeed, if one tries to fix $\sigma_i'$ in $E_{\text{NDR}}'$ according to the reduction of \eqref{eq:wrongreduce}
via Eq.~\eqref{eq:NDRreduce}, one no longer sees the explicit cancellation of the evanescent-to-physical
mixing.


At this point, we wish to stress that this prescription only applies to non-Fierz cases. It is unclear to the authors whether such a recursive method can 
be extended to the more general ``Fierz-like'' case in Eq.~\eqref{eq:nonFierz}, and for the time being must be treated on a case-by-case 
basis\footnote{For a discussion on Fierz-evanescent operators in the context of one-loop Fierz identities we refer to \cite{Aebischer:2022aze,Aebischer:2022rxf,Aebischer:2023djt}.}.

In the following two subsections we will apply this method to two examples to reproduce known results from the literature, namely for $|\Delta F| = 1$ and $|\Delta F| = 2$ two-loop ADMs.

\subsection{Example I: Two-Loop QCD ADM for $|\Delta F| = 1$}

As a first example we begin by considering the charged-current weak decay governed by
the effective Hamiltonian
\begin{equation}\label{eq:HDF1}
	\mathcal{H}_{\Delta F = 1} \supset \frac{4 G_F}{\sqrt{2}}\sum_{k,\ell = u, c} V_{k r}^* V_{\ell d}
		\Big(C_+ Q^{k\ell}_+ + C_- Q^{k\ell}_-\Big)\,,
\end{equation}
where $r = s, b$ for $|\Delta S| = 1$ and $|\Delta B| = 1$ processes, respectively and
\begin{equation}
	Q_{\pm}^{qq'} = \frac{1}{2}\Big(\big(\bar{r}^\alpha_L\gamma^\mu q_L^\alpha\big)\big(\bar{q}^{\prime \beta}_L \gamma_\mu d^\beta_L\big)
		\pm \big(\bar{r}^\alpha_L\gamma^\mu q^\beta_L\big)\big(\bar{q}^{\prime \beta}_L \gamma_\mu d^\alpha_L\big)\Big)\,.
\end{equation}

Inserting these operators into one-loop diagrams with a single gluon generates four non-trivial structures to which we
assign the replacements
\begin{equation}\label{eq:4qprescrip}
	\begin{split}
		\big(\gamma^\mu \gamma^\nu \gamma^\alpha P_L \gamma_\nu \gamma_\mu\big)\otimes\big(\gamma_\alpha P_L\big)
		\to& (4 - \epsilon \sigma_F)\big(\gamma^\alpha P_L\big)\otimes\big(\gamma_\alpha P_L\big)\,, \\[5pt]
		\big(\gamma^\alpha P_L \gamma^\mu \gamma^\nu\big)\otimes \big(\gamma_\nu \gamma_\mu \gamma_\alpha P_L\big)
		\to& (4 - \epsilon \sigma_{V1})\big(\gamma^\alpha P_L\big)\otimes\big(\gamma_\alpha P_L\big)\,, \\[5pt]
		\big(\gamma^\mu \gamma^\nu \gamma^\alpha P_L\big)\otimes\big(\gamma_\nu \gamma_\mu \gamma_\alpha P_L\big)
		\to& (- 8 - \epsilon \sigma_{V2})\big(\gamma^\alpha P_L\big)\otimes\big(\gamma_\alpha P_L\big)\,, \\[5pt]
		\big(\gamma^\alpha P_L\gamma^\mu \gamma^\nu\big)\otimes\big(\gamma_\alpha P_L \gamma_\nu \gamma_\mu\big)
		\to& (- 8 - \epsilon \sigma_{V3})\big(\gamma^\alpha P_L\big)\otimes\big(\gamma_\alpha P_L\big)\,.
	\end{split}
\end{equation}
Furthermore, for future simplicity, we choose $\sigma_F = \sigma_{V1}$. In this basis, the one-loop QCD ADM is diagonal.
It can be found, for example, in Ref.s~\cite{Herrlich:1996vf,Buchalla:1995vs}, and since it is scheme-independent, we will trivially
reproduce the same result.

At two-loops, all $1/\epsilon^2$ poles reduce using only the relations given in Eq.~\eqref{eq:4qprescrip}, as they must
for the cancellation of subdivergences. At $O(1/\epsilon)$, additional irreducible structures arise. These structures can be treated in an identical way, but for an NLO ADM computation, only the four-dimensional piece is relevant, so we do not give the explicit relations.

We separate the results by their scheme-dependence as
\begin{equation}
	\gamma^{(1)}=\gamma^{(1),\text{S.I.}} + \sigma_{V1}\gamma^{(1),V1} + \sigma_{V2}\gamma^{(1),V2} + \sigma_{V3}\gamma^{(1),V3} \,,
\end{equation}
and we find (fixing the number of colors $N_c = 3$ and number of active flavors $f = 5$)
\begin{equation}\label{eq:dF1APP}
	\begin{split}
		\gamma^{(1),\text{S.I.}} = \begin{pmatrix}
			\frac{877}{9} &&& 0 \\[5pt]
			0 &&& \frac{202}{9}
		\end{pmatrix} \quad &, \quad
		\gamma^{(1),V1} = \begin{pmatrix}
			-\frac{115}{9} &&& 0 \\[5pt]
			0 &&& -\frac{46}{9}
		\end{pmatrix}\,,\\[10pt]
		\gamma^{(1),V2} = \gamma^{(1), V3} &= \begin{pmatrix}
			-\frac{23}{18} &&& 0 \\[5pt]
			0 &&& \frac{23}{9}
		\end{pmatrix}\,.
	\end{split}
\end{equation}

In Ref.~\cite{Herrlich:1996vf}, the same ADM is computed in the NDR scheme using the evanescent operator
\begin{equation}\label{eq:NDRevan}
	E = \big(\gamma^\mu \gamma^\nu \gamma^\alpha P_L\big)\otimes\big(\gamma_\mu \gamma_\nu \gamma_\alpha P_L\big)
		- (16 - a\epsilon)\big(\gamma^\alpha P_L\big)\otimes\big(\gamma_\alpha P_L\big)\,,
\end{equation}
from which, one finds
\begin{equation}\label{eq:dF1NDR}
	\gamma_{\text{NDR}}^{(1)} = \begin{pmatrix}
		-\frac{365}{9} + \frac{161}{18}a &&& \frac{20}{3} - \frac{5}{3} a \\[5pt]
		\frac{82}{3} - \frac{41}{6}a &&& -\frac{74}{9} - \frac{23}{9}a
	\end{pmatrix}\,,
\end{equation}
which reduces to that given in \cite{Herrlich:1996vf} for the standard choice of $a = 4$. After performing a change of 
renormalization scheme (see e.g. Ref.s~\cite{Gorbahn:2004my,Chetyrkin:1997gb,Aebischer:2023djt}) from Eq.s~\eqref{eq:4qprescrip} and \eqref{eq:NDRevan} and
applying the resulting transformation to Eq.~\eqref{eq:dF1APP}, we exactly reproduce Eq.~\eqref{eq:dF1NDR}.

Here, we wish to emphasize a few critical points. First, we see that in our scheme, the $|\Delta F| = 1$ QCD mixing
remains diagonal for \textit{any choice of scheme constants}. This is in contrast to the case of NDR which is only diagonal for
a particular choice of scheme. 

To understand why this occurs, we first note that the $Q_\pm$ operators are self-Fierz and anti-self-Fierz conjugate operators, respectively.
Consider the Fierz-evanescent operator corresponding to a $V-A$ operator insertion
\begin{equation}
	E_F = (\gamma^\mu P_L)_{i\ell}(\gamma_\mu P_L)_{kj} - (1 - \epsilon\sigma_{\mathcal{F}})
		(\gamma^\mu P_L)_{ij}(\gamma_\mu P_L)_{k\ell}\,,
\end{equation}
where $i,j,k,\ell$ are Dirac indices. After insertion into a one-loop diagram, we find
\begin{equation}
	\begin{split}
		\expval{E_F}^{(1)}&\propto (\Gamma_1)_{mi}(\Gamma_2)_{jn}(\Gamma_3)_{rk}(\Gamma_4)_{\ell q}\big[
			(\gamma^\alpha P_L)_{i\ell}(\gamma_\alpha P_L)_{kj} - (1 - \epsilon\sigma_{\mathcal{F}})(\gamma^\alpha P_L)_{ij}(\gamma_\alpha P_L)_{k\ell}\big]\\[0.5em]
		&= (\Gamma_1\gamma^\alpha P_L\Gamma_4)\otimes(\Gamma_3\gamma_\alpha P_L\Gamma_2) - (1 - \epsilon\sigma_{\mathcal{F}})
			(\Gamma_1\gamma^\alpha P_L\Gamma_2)\otimes(\Gamma_3\gamma_\alpha P_L\Gamma_4)\,.
	\end{split}
\end{equation}
Setting $\Gamma_1 = \gamma^\mu\gamma^\nu$, $\Gamma_2 = \mathbb{1}$, $\Gamma_3 = \gamma_\nu\gamma_\mu$, and $\Gamma_4 = \mathbb{1}$, both terms
in the one-loop insertion of $E_F$ reduce the same way using the third relation in Eq.~\eqref{eq:4qprescrip} and this contribution vanishes for $\sigma_{\mathcal{F}} = 0$. 
The same is true for $\Gamma_2 = \gamma^\mu\gamma^\nu$, $\Gamma_4 = \gamma_\nu\gamma_\mu$, $\Gamma_1 = \Gamma_3 = \mathbb{1}$, using the fourth reduction in Eq.~\eqref{eq:4qprescrip}. However,
for $\Gamma_1 = \gamma^\mu\gamma^\nu$, $\Gamma_4 = \gamma_\nu\gamma_\mu$, $\Gamma_2 = \Gamma_3 = \mathbb{1}$, the two terms reduce using the first and second lines of Eq.~\eqref{eq:4qprescrip},
respectively, and the one-loop Fierz-evanescent insertion only vanishes for the specific choice $\sigma_F = \sigma_{V1}$. Hence, using this scheme choice preserves tree-level Fierz 
identities and consequently the diagonality of the ADM.

Next, we reiterate that we only needed to consider physical operators and we worked in a pure
$\ol{\text{MS}}$ scheme, requiring no finite subtractions. Furthermore, we did not need to fix our physical basis
in order to eliminate evanescent operators; the relations appearing in Eq.~\eqref{eq:4qprescrip} guarantee the preservation
of four-dimensional Dirac relations when inserted into loop diagrams and recursively reduced. Finally, since we reproduce the NDR result 
using only a standard transformation of renormalization scheme, our method is equivalent to the computation in NDR up to
the unphysical choice of scheme.

\subsection{Example II: Two-Loop QCD ADM for $|\Delta S| = 2$}

In this section, we show that the proposed scheme can also be applied to other non-trivial scenarios aside from
four-quark operators with gauge boson loops. In this case, we consider the $|\Delta S| = 2$ Hamiltonian
\begin{equation}
	\mathcal{H}_{\Delta S = 2} = - \frac{G_F^2 M_W^2}{4\pi^2}\lambda_t^2 C_{S2} Q_{S2}
		+ 8 G_F^2 \lambda_u \lambda_t \tilde{C}_7 \tilde{Q}_7\,,
\end{equation}
where
\begin{equation}
	Q_{S2} = \big(\bar{s}_L^\alpha \gamma^\mu d_L^\alpha\big)\big(\bar{s}_L^\beta \gamma_\mu d_L^\beta\big)\,, \quad
	\tilde{Q}_7 = \frac{m_c^2}{4\pi\alpha_s}Q_{S2}\,,
\end{equation}
and $\lambda_i = V^*_{is}V_{id}$. Here, we have used the $u-t$ unitary basis \cite{Brod:2019rzc} instead of the $c-t$ basis used in 
Ref.~\cite{Herrlich:1996vf}. Due to the large suppression from $\lambda_t^2$, the dimension-eight operator $\tilde{Q}_7$ gives a similar
contribution as that from $Q_{S2}$ despite the additional $m_c^2/M_W^2$ suppression. The WC $\tilde{C}_7$ obtains
no matching conditions, but is instead generated by RG running from double-insertions of the $|\Delta S| = 1$ operators in Eq.~\eqref{eq:HDF1} 
(we neglect penguin operator insertions and only focus on the charged-current subspace).

In our scheme, we use the same relations as given in Eq.~\eqref{eq:4qprescrip}, and the one-loop double insertion gives one additional
relation
\begin{equation}
	\big(\gamma^\alpha P_L \gamma^\mu \gamma^\beta P_L\big)\otimes \big(\gamma_\beta P_L \gamma_\mu \gamma_\alpha P_L\big) =
		(4 - \epsilon \sigma_{V4})\big(\gamma^\alpha P_L\big)\otimes\big(\gamma_\alpha P_L\big)\,.
\end{equation}
In principle, we also need the Fierz-evanescent operators
\begin{equation}\label{eq:fierzevan}
	\begin{split}
		F_{S2} =& \big(\bar{s}_L^\alpha \gamma^\mu d_L^\beta\big)\big(\bar{s}_L^\beta \gamma_\mu d_L^\alpha\big)
			- (1 - \epsilon F_1) Q_{S2}\,, \\[5pt]
		\tilde{F}_7 =& \frac{m_c^2}{4\pi\alpha_s}
			\big(\bar{s}_L^\alpha \gamma^\mu d_L^\beta\big)\big(\bar{s}_L^\beta \gamma_\mu d_L^\alpha\big)
			- (1 - \epsilon \tilde{F}_1) \tilde{Q}_7\,,
	\end{split}
\end{equation}
however, the ($N_c$- and $f$-independent) choice of $\sigma_F = \sigma_{V1}$ along with $F_1 = \tilde{F}_1 = 0$ leads to
vanishing evanescent-to-physical mixing at this order as discussed in the previous section, so the evanescent operators may be ignored with this choice
and we used simple tree-level Fierz relations.

Again, at two-loop the $1/\epsilon^2$ poles reduce using only the one-loop relations, and more relations must be introduced
for the $1/\epsilon$ poles. The single-insertion mixing can be immediately extracted from the
``$++$'' component of Eq.~\eqref{eq:dF1APP} since $Q_{S2}$ is equal to its Fierz conjugate in this scheme.

The two-loop anomalous dimension tensor computed from double insertions of $|\Delta S| = 1$ operators with gluon loops is given by
\begin{equation}
	\begin{split}
		\hat{\gamma}^{(1),\text{S.I.}}_7 = \begin{pmatrix}
			14 &&& -\frac{5}{3} \\[5pt]
			-\frac{5}{3} &&& \frac{26}{3}
		\end{pmatrix}&, \quad
		\hat{\gamma}^{(1),V1}_7 = \begin{pmatrix}
			-\frac{5}{2} &&& \frac{1}{3} \\[5pt]
			\frac{1}{3} &&& \frac{1}{6}
		\end{pmatrix}\,,\\[10pt]
		\hat{\gamma}^{(1),V2}_7 = \hat{\gamma}^{(1),V3}_7 = \begin{pmatrix}
			-\frac{1}{4} &&& -\frac{1}{6} \\[5pt]
			-\frac{1}{6} &&& \frac{5}{12}
		\end{pmatrix}&, \quad
		\hat{\gamma}^{(1),V4}_7 = \begin{pmatrix}
			\frac{9}{2} &&& - 3 \\[5pt]
			- 3 &&& \frac{9}{2}
		\end{pmatrix}\,.
	\end{split}
\end{equation}

The NDR results of the $|\Delta S| = 2$ mixing can, in principle, be extracted from Ref.~\cite{Herrlich:1996vf} using the basis change 
in Ref.~\cite{Brod:2019rzc} along with a partial conversion from the $Q_1-Q_2$ basis to the $Q_+-Q_-$ basis. We instead re-computed the
results to find (again, setting $F_1 = \tilde{F}_1 = 0$ in Eq.~\eqref{eq:fierzevan})
\begin{equation}
	\gamma^{(1)}_{S2;\text{NDR}} = -\frac{227}{9} + \frac{46}{9}a\,,
\end{equation}
for the single-insertion mixing and
\begin{equation}
	\hat{\gamma}_{7;\text{NDR}}^{(1)} = \begin{pmatrix}
		34 - a &&& -\frac{89}{3} + \frac{5}{3}a \\[5pt]
		-\frac{89}{3} + \frac{5}{3} a &&& \frac{194}{3} - \frac{14}{3} a
	\end{pmatrix}\,,
\end{equation}
for the anomalous dimension tensor. After converting renormalization schemes, we find exact agreement between the two methods.

\section{Conclusions}\label{sec:concl}

In this article we have introduced a novel prescription to treat effects from evanescent operators. Besides other possible simplifications the method ensures the vanishing of evanescent-to-physical mixing \textit{a priori} which simplifies calculations significantly. Instead of having to remove such contributions by introducing finite counterterms, a simple set of reduction rules is introduced. This set is used recursively in order to map all occurring Dirac structures back to the physical basis, without the need to insert evanescent operators into loop diagrams. The rules only have to be derived once and can then be used in further calculations since the relations are process-independent. 
The approach is independent of the treatment of $\gamma_5$ and can therefore be used in combination with different treatments of the Dirac algebra in the mass-dimension-four sector of the
theory. 

We illustrate the usefulness of this approach in two examples from the literature, where we reproduce known results for $|\Delta F| = 1$ and $|\Delta F| = 2$ two-loop QCD ADMs. The same reduction rules applied in both processes, underlining the simplicity of the approach. 

The results found in this article can be applied for instance to calculations of two-loop ADMs in the Standard Model Effective Theory (SMEFT) or the Weak Effective Theory (WET). The matching between these two theories is known at the tree-level \cite{Aebischer:2015fzz,Jenkins:2017jig} and at the one-loop level \cite{Dekens:2019ept}. Furthermore, the RG running is known at the one-loop level in the WET \cite{Jenkins:2017dyc,Aebischer:2017gaw} and SMEFT \cite{Alonso:2013hga, Jenkins:2013zja, Jenkins:2013wua} and since recently, the complete evanescent basis for the WET has been derived in the HV scheme in \cite{Naterop:2023dek}. Hence, the two-loop running is the only missing piece for a complete NLO analysis of the SMEFT and WET. With the approach presented in this article this computation would be simplified considerably compared to conventional methods, since evanescent insertions together with the renormalization of EVs can be neglected.

We have also shown in an explicit case that a careful treatment of the proposed scheme can in some cases also ensure the vanishing of the mixing of Fierz-evanescent operators into the
physical basis, but it is at this point unclear whether this is a general feature or only applicable to the considered cases. In the case where this method does not generalize, one must still
consider loop-insertions of Fierz-evanescent operators, or use one-loop Fierz identities \cite{Aebischer:2022aze,Aebischer:2022rxf,Aebischer:2023djt}. Furthermore, a matching computation onto 
this scheme might involve rather complicated scheme transformations. Finally, a generalization of this approach to higher orders or to include also non-internal structures would be desirable. 
We leave such studies for the future.

Finally, we note that the presented prescription method for performing NLO calculations is well suited for automation, due to its simple and recursive character. 
It would therefore be very interesting to implement this approach in order to make it accessible for the community.


\section*{Acknowledgements}
\addcontentsline{toc}{section}{\numberline{}Acknowledgements}

J.\ A.\,, M.\ P.\ and Z.\ P.\ acknowledge  financial  support  from  the  European  Research  Council  (ERC)  under the European Union's Horizon 2020 research and innovation programme under grant agreement 833280 (FLAY), and from the Swiss National Science Foundation (SNF) under contract 200020-204428. We also would like to thank Felix Wilsch for fruitful conversations regarding the treatment of Fierz-evanescent operators.


\clearpage

\appendix

\section{Two-loop contributions}\label{app:2Lex}

\begin{figure}[H]
\hspace{-1cm}
	\begin{subfigure}[c]{.3\textwidth}
		\centering
		\includegraphics[width=.6\linewidth]{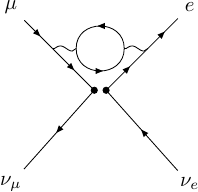}
		\caption{}
		\label{fig:sfig1}
	  \end{subfigure}
\hspace{-1cm}
	\begin{subfigure}[c]{.3\textwidth}
		\centering
		\includegraphics[width=.6\linewidth]{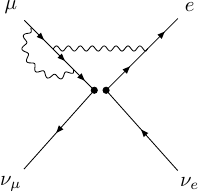}
		\caption{}
		\label{fig:sfig2}
	  \end{subfigure}
\hspace{-1cm}
	\begin{subfigure}[c]{.3\textwidth}
		\centering
		\includegraphics[width=.6\linewidth]{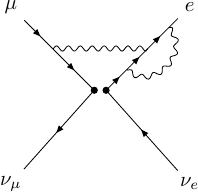}
		\caption{}
		\label{fig:sfig3}
	  \end{subfigure}
\hspace{-1cm}
	\begin{subfigure}[c]{.3\textwidth}
		\centering
		\includegraphics[width=.6\linewidth]{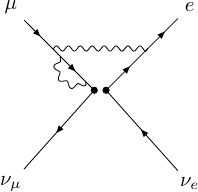}
		\caption{}
		\label{fig:sfig4}
	  \end{subfigure}
\hspace{1cm}
	\begin{subfigure}[c]{.3\textwidth}
	  \centering
	  \includegraphics[width=.6\linewidth]{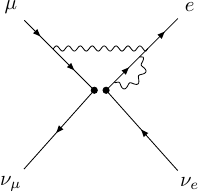}
	  \caption{}
	  \label{fig:sfig5}
	\end{subfigure}
	\begin{subfigure}[c]{.3\textwidth}
		\centering
		\includegraphics[width=.6\linewidth]{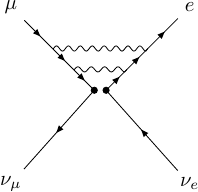}
		\caption{}
		\label{fig:sfig7}
	  \end{subfigure}
	  \begin{subfigure}[c]{.3\textwidth}
		\centering
		\includegraphics[width=.6\linewidth]{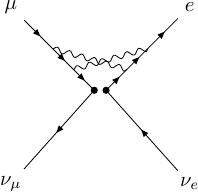}
		\caption{}
		\label{fig:sfig8}
	  \end{subfigure}%
	\caption{Genuine two-loop diagrams relevant for the muon decay in the EFT, discussed in Sec.~\ref{sec:example}.}
	\label{fig:2L}
	\end{figure}

\begin{figure}[H]
	\begin{subfigure}[c]{.3\textwidth}
		\centering
		\includegraphics[width=.6\linewidth]{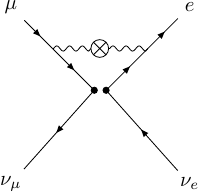}
		\caption{}
		\label{fig:sfig1}
	  \end{subfigure}
	\begin{subfigure}[c]{.3\textwidth}
		\centering
		\includegraphics[width=.6\linewidth]{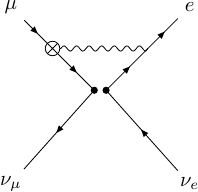}
		\caption{}
		\label{fig:sfig2}
	  \end{subfigure}
	\begin{subfigure}[c]{.3\textwidth}
		\centering
		\includegraphics[width=.6\linewidth]{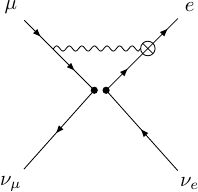}
		\caption{}
		\label{fig:sfig3}
	  \end{subfigure}
\centering
	\begin{subfigure}[c]{.3\textwidth}
		\centering
		\includegraphics[width=.6\linewidth]{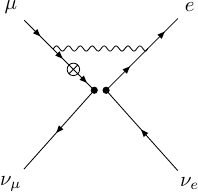}
		\caption{}
		\label{fig:sfig4}
	  \end{subfigure}
	\begin{subfigure}[c]{.3\textwidth}
	  \centering
	  \includegraphics[width=.6\linewidth]{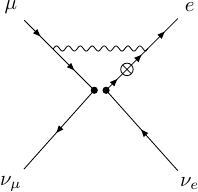}
	  \caption{}
	  \label{fig:sfig5}
	\end{subfigure}%
	\caption{Counterterm diagrams for the muon decay in the EFT, discussed in Sec.~\ref{sec:example}.}
	\label{fig:CT}
	\end{figure}

In this appendix we collect all relevant information as well as the two-loop diagrams that are necessary for the computation in Sec.~\ref{sec:example}. The genuine two-loop diagrams are depicted in Fig.~\ref{fig:2L}. In Tab.~\ref{tab:2loop}, the $1/\epsilon^2$ and $1/\epsilon$ poles of the individual diagrams are collected for a general operator insertion of the form
\begin{equation}
	(\Gamma_1)\otimes(\Gamma_2)\,,
\end{equation}
with a WC $C_{12}$. The results are given in units of $-iC_{12}(\alpha/(4\pi))^2$. The different entries correspond to the diagrams (a)-(g) depicted in Fig.~\ref{fig:2L}. In Tab.~\ref{tab:CTs} the counterterm insertions are collected, and the corresponding diagrams (a)-(e) are shown in Fig.~\ref{fig:CT}. The one-loop counterterms in the $\overline{\text{MS}}$-scheme and general $R_\xi$ gauge are given by
\begin{equation}
	\begin{split}
		&\delta Z_\ell^{(1;1)} = -\xi\,Q_\ell^2\,, \qquad \delta Z_A^{(1;1)} = -\frac{4}{3}\sum_\ell Q_\ell^2 n_\ell\,, \\
		& \delta Z_e^{(1;1)} = \frac{4}{6}\sum_\ell Q_\ell^2 n_\ell\,, \qquad \delta Z_{\chi^2}^{(1;1)} = -4\sum_\ell Q_\ell^2 n_\ell\,,
	\end{split}
\end{equation}
where $n_\ell$ is the number of charged fermions with charge $Q_\ell$ and $\delta Z_{\chi^2}$ is the counterterm corresponding to
the artificial IR regulator mass given to the photon.

At two-loops, only the renormalization constants for the external fields are needed, given by
\begin{equation}
	\delta Z_\ell^{(2;2)} = \frac{\xi^2}{2}Q_\ell^4\,, \qquad \delta Z_\ell^{(2;1)} = \frac{3}{4} Q_\ell^4 + Q_\ell^2\sum_{\ell'} Q_{\ell'}^2 n_{\ell'}\,.
\end{equation}
For simplicity, the results in Tab.s~\ref{tab:2loop}-\ref{tab:CTs} are presented in the Feynman gauge, whereas the computation was performed in the general $R_\xi$ gauge.

\begin{table}[H]
	\resizebox{\columnwidth}{!}{%
	\begin{tabular}{lll}
		\cline{1-3}
		\multicolumn{1}{|l|}{} & \multicolumn{1}{c|}{1/$\epsilon^2$ pole} & \multicolumn{1}{c|}{1/$\epsilon$ pole} \\
		\cline{1-3}
		\multicolumn{1}{|c|}{Fermion Loop $(n_f Q_f^2)$} & \multicolumn{1}{c|}{\makecell{
			$\frac{1}{72}\big(\Gamma_1\gamma^\mu\gamma^\nu\big)\otimes\big(\gamma^\alpha\gamma_\mu\Gamma_2\big)
				\text{Tr}\big[\gamma_\nu \gamma^\beta \gamma_\alpha \gamma_\beta\big]$ \\
			$\frac{1}{288}\big(\Gamma_1\gamma^\mu\gamma^\nu\big)\otimes\big(\gamma^\alpha\gamma^\beta\Gamma_2\big)
				\text{Tr}\big[\gamma_\nu \gamma_\mu \gamma_\alpha \gamma_\beta\big]$ \\			
			$\frac{1}{288}\big(\Gamma_1\gamma^\mu\gamma^\nu\big)\otimes\big(\gamma^\alpha\gamma^\beta\Gamma_2\big)
				\text{Tr}\big[\gamma_\nu \gamma_\beta \gamma_\alpha \gamma_\mu\big]$}}
						   & \multicolumn{1}{c|}{\makecell{
			$\frac{5}{108}\big(\Gamma_1\gamma^\mu\gamma^\nu\big)\otimes\big(\gamma^\alpha\gamma_\mu\Gamma_2\big)
				\text{Tr}\big[\gamma_\nu \gamma^\beta \gamma_\alpha \gamma_\beta\big]$ \\
			$- \frac{1}{1728}\big(\Gamma_1\gamma^\mu\gamma^\nu\big)\otimes\big(\gamma^\alpha\gamma^\beta\Gamma_2\big)
				\text{Tr}\big[\gamma_\nu \gamma_\mu \gamma_\alpha \gamma_\beta\big]$ \\			
			$- \frac{1}{1728}\big(\Gamma_1\gamma^\mu\gamma^\nu\big)\otimes\big(\gamma^\alpha\gamma^\beta\Gamma_2\big)
				\text{Tr}\big[\gamma_\nu \gamma_\beta \gamma_\alpha \gamma_\mu\big]$}} \\
		\cline{1-3}
		\multicolumn{1}{|c|}{Vert. Corr. ($\mu$)} & \multicolumn{1}{c|}{\makecell{
			$\frac{1}{32}(\Gamma_1\gamma^\mu \gamma^\alpha \gamma^\beta \gamma^\nu \gamma_\beta \gamma_\alpha)
				\otimes(\gamma_\nu \gamma_\mu\Gamma_2)$}}
						   & \multicolumn{1}{c|}{\makecell{
			$\frac{1}{192}(\Gamma_1\gamma^\mu \gamma^\alpha \gamma^\beta \gamma^\nu \gamma_\beta \gamma_\alpha)
				\otimes(\gamma_\nu \gamma_\mu\Gamma_2)$ \\
			$-\frac{1}{96}(\Gamma_1\gamma^\alpha \gamma^\beta \gamma^\mu \gamma^\nu \gamma_\alpha\gamma_\beta)
				\otimes(\gamma_\nu \gamma_\mu \Gamma_2)$ \\
			$-\frac{1}{96}(\Gamma_1\gamma^\alpha \gamma^\beta \gamma_\alpha \gamma^\mu \gamma^\nu \gamma_\beta)
				\otimes(\gamma_\mu \gamma_\nu \Gamma_2)$}} \\
		\cline{1-3}
		\multicolumn{1}{|c|}{Vert. Corr. ($\bar{e}$)} & \multicolumn{1}{c|}{\makecell{
			$\frac{1}{32}(\Gamma_1\gamma^\mu \gamma^\nu)
				\otimes(\gamma^\alpha\gamma^\beta\gamma_\nu\gamma_\beta\gamma_\alpha\gamma_\mu\Gamma_2)$}}
						   & \multicolumn{1}{c|}{\makecell{
			$\frac{1}{192}(\Gamma_1\gamma^\mu \gamma^\nu)
				\otimes(\gamma^\alpha\gamma^\beta\gamma_\nu\gamma_\beta\gamma_\alpha\gamma_\mu\Gamma_2)$ \\
			$-\frac{1}{96}(\Gamma_1\gamma^\mu \gamma^\nu)
				\otimes(\gamma^\alpha\gamma^\beta\gamma_\nu\gamma_\mu\gamma_\alpha\gamma_\beta \Gamma_2)$ \\
			$-\frac{1}{96}(\Gamma_1\gamma^\mu\gamma^\nu)
				\otimes(\gamma^\alpha\gamma_\mu \gamma_\nu\gamma^\beta\gamma_\alpha\gamma_\beta\Gamma_2)$}} \\
		\cline{1-3}
		\multicolumn{1}{|c|}{Fermion SE ($\mu$)} & \multicolumn{1}{c|}{\makecell{
			$\frac{1}{96}(\Gamma_1 \gamma^\alpha \gamma^\beta \gamma_\alpha \gamma_\beta \gamma^\mu \gamma^\nu)
				\otimes(\gamma_\nu\gamma_\mu \Gamma_2)$ \\
			$+\frac{1}{96}(\Gamma_1 \gamma^\alpha \gamma^\beta\gamma^\mu\gamma_\beta\gamma_\alpha\gamma^\nu)
				\otimes(\gamma_\nu\gamma_\mu \Gamma_2)$ \\
			$+\frac{1}{96}(\Gamma_1 \gamma^\mu\gamma^\alpha\gamma^\beta\gamma_\alpha\gamma_\beta\gamma^\nu)
				\otimes(\gamma_\nu\gamma_\mu \Gamma_2)$}}
						   & \multicolumn{1}{c|}{\makecell{
			$\frac{1}{576}(\Gamma_1 \gamma^\alpha \gamma^\beta \gamma_\alpha \gamma_\beta \gamma^\mu \gamma^\nu)
				\otimes(\gamma_\nu\gamma_\mu \Gamma_2)$ \\
			$+\frac{1}{576}(\Gamma_1 \gamma^\alpha \gamma^\beta\gamma^\mu\gamma_\beta\gamma_\alpha\gamma^\nu)
				\otimes(\gamma_\nu\gamma_\mu \Gamma_2)$ \\
			$+\frac{1}{576}(\Gamma_1 \gamma^\mu\gamma^\alpha\gamma^\beta\gamma_\alpha\gamma_\beta\gamma^\nu)
				\otimes(\gamma_\nu\gamma_\mu \Gamma_2)$}} \\
		\cline{1-3}
		\multicolumn{1}{|c|}{Fermion SE ($\bar{e}$)} & \multicolumn{1}{c|}{\makecell{
			$\frac{1}{96}(\Gamma_1\gamma^\mu\gamma^\nu)
				\otimes(\gamma_\nu\gamma_\mu\gamma^\alpha\gamma^\beta\gamma_\alpha\gamma_\beta\Gamma_2)$ \\
			$+\frac{1}{96}(\Gamma_1\gamma^\mu\gamma^\nu)
				\otimes(\gamma_\nu\gamma^\alpha\gamma^\beta\gamma_\mu\gamma_\beta\gamma_\alpha\Gamma_2)$ \\
			$+\frac{1}{96}(\Gamma_1\gamma^\mu\gamma^\nu)
				\otimes(\gamma_\nu\gamma^\alpha\gamma^\beta\gamma_\alpha\gamma_\beta\gamma_\mu\Gamma_2)$}}
						   & \multicolumn{1}{c|}{\makecell{
			$\frac{1}{576}(\Gamma_1\gamma^\mu\gamma^\nu)
				\otimes(\gamma_\nu\gamma_\mu\gamma^\alpha\gamma^\beta\gamma_\alpha\gamma_\beta\Gamma_2)$ \\
			$+\frac{1}{576}(\Gamma_1\gamma^\mu\gamma^\nu)
				\otimes(\gamma_\nu\gamma^\alpha\gamma^\beta\gamma_\mu\gamma_\beta\gamma_\alpha\Gamma_2)$ \\
			$+\frac{1}{576}(\Gamma_1\gamma^\mu\gamma^\nu)
				\otimes(\gamma_\nu\gamma^\alpha\gamma^\beta\gamma_\alpha\gamma_\beta\gamma_\mu\Gamma_2)$}} \\
		\cline{1-3}
		\multicolumn{1}{|c|}{Ladder} & \multicolumn{1}{c|}{\makecell{
			$\frac{1}{32}(\Gamma_1\gamma^\mu \gamma^\nu \gamma^\alpha \gamma^\beta)
				\otimes(\gamma_\beta\gamma_\alpha\gamma_\nu\gamma_\mu\Gamma_2)$}}
						   & \multicolumn{1}{c|}{\makecell{
			$\frac{5}{192}(\Gamma_1\gamma^\mu \gamma^\nu \gamma^\alpha \gamma^\beta)
				\otimes(\gamma_\beta\gamma_\alpha\gamma_\nu\gamma_\mu\Gamma_2)$ \\
			$+\frac{1}{96}(\Gamma_1 \gamma^\alpha \gamma^\mu \gamma_\alpha \gamma^\nu)
				\otimes(\gamma_\nu\gamma^\beta\gamma_\mu\gamma_\beta\Gamma_2)$ \\
			$+\frac{1}{96}(\Gamma_1\gamma^\mu\gamma^\nu\gamma^\alpha\gamma^\beta)
				\otimes(\gamma_\beta\gamma_\mu\gamma_\nu\gamma_\alpha\Gamma_2)$}} \\
		\cline{1-3}
		\multicolumn{1}{|c|}{Crossed Ladder} & \multicolumn{1}{c|}{\makecell{
			0 }}
						   & \multicolumn{1}{c|}{\makecell{
			$-\frac{1}{24}(\Gamma_1\gamma^\mu\gamma^\nu\gamma^\alpha\gamma^\beta)
				\otimes(\gamma_\nu\gamma_\alpha\gamma_\beta\gamma_\mu\Gamma_2)$ \\
			$+\frac{1}{48}(\Gamma_1\gamma^\alpha\gamma^\mu\gamma_\alpha\gamma^\nu)
				\otimes(\gamma_\mu\gamma^\beta\gamma_\nu\gamma_\beta\Gamma_2)$ \\
			$+\frac{1}{48}(\Gamma_1\gamma^\mu\gamma^\nu\gamma^\alpha\gamma^\beta)
				\otimes(\gamma_\nu\gamma_\mu\gamma_\beta\gamma_\alpha\Gamma_2)$}} \\
		\cline{1-3}
	\end{tabular}
	}
	\caption{Two-loop results with an insertion of operator with structure $(\Gamma_1)~\otimes~(\Gamma_2)$. The results
	are presented immediately after loop integration with no reduction of Dirac algebra. All results are multiplied by
	$-iC_{12}(\alpha/(4\pi))^2$ and non-local divergences are neglected as they are equal to twice the $1/\epsilon^2$ poles. Furthermore, we set $Q_\mu=Q_e=-1$.}
	\label{tab:2loop}
\end{table}

\begin{table}[H]
	\centering
	\resizebox{\columnwidth}{!}{%
	\begin{tabular}{lll}
		\cline{1-3}
		\multicolumn{1}{|l|}{} & \multicolumn{1}{c|}{1/$\epsilon^2$ pole} & \multicolumn{1}{c|}{1/$\epsilon$ pole} \\
		\cline{1-3}
		\multicolumn{1}{|c|}{Photon Field/Mass CT ($n_f Q_f^2$)} & \multicolumn{1}{c|}{\makecell{
			$\frac{5}{18}(\Gamma_1\gamma^\mu\gamma^\nu)\otimes(\gamma_\nu\gamma_\mu\Gamma_2)$ \\
			$-\frac{1}{18}(\Gamma_1\gamma^\mu\gamma^\nu)\otimes(\gamma_\mu\gamma_\nu\Gamma_2)$ \\
			$-\frac{1}{18}(\Gamma_1\gamma^\mu\gamma_\mu)\otimes(\gamma^\nu\gamma_\nu\Gamma_2)$
			}}
						   & \multicolumn{1}{c|}{\makecell{
			$\frac{2}{9}(\Gamma_1\gamma^\mu\gamma^\nu)\otimes(\gamma_\nu\gamma_\mu\Gamma_2)$
			}} \\
		\cline{1-3}
		\multicolumn{1}{|c|}{Vertex CT ($\mu$)} & \multicolumn{1}{c|}{\makecell{
			$-\frac{1}{4}(\Gamma_1\gamma^\mu\gamma^\nu)\otimes(\gamma_\nu\gamma_\mu\Gamma_2)$
			}}
						   & \multicolumn{1}{c|}{\makecell{
			$ 0 $}} \\
		\cline{1-3}
		\multicolumn{1}{|c|}{Vertex CT ($\bar{e}$)} & \multicolumn{1}{c|}{\makecell{
			$-\frac{1}{4}(\Gamma_1\gamma^\mu\gamma^\nu)\otimes(\gamma_\nu\gamma_\mu\Gamma_2)$
			}}
						   & \multicolumn{1}{c|}{\makecell{
			$ 0 $}} \\
		\cline{1-3}
		\multicolumn{1}{|c|}{Fermion Field CT ($\mu$)} & \multicolumn{1}{c|}{\makecell{
			$\frac{1}{24}(\Gamma_1\gamma^\alpha\gamma_\alpha\gamma^\mu\gamma^\nu)\otimes
				(\gamma_\nu\gamma_\mu\Gamma_2)$ \\
			$+ \frac{1}{24}(\Gamma_1\gamma^\alpha\gamma^\mu\gamma_\alpha\gamma^\nu)\otimes
				(\gamma_\nu\gamma_\mu\Gamma_2)$ \\
			$+ \frac{1}{24}(\Gamma_1\gamma^\mu\gamma^\alpha\gamma_\alpha\gamma^\nu)\otimes
				(\gamma_\nu\gamma_\mu\Gamma_2)$
			}}
						   & \multicolumn{1}{c|}{\makecell{
			$ 0 $}} \\
		\cline{1-3}
		\multicolumn{1}{|c|}{Fermion Field CT ($\bar{e}$)} & \multicolumn{1}{c|}{\makecell{
			$\frac{1}{24}(\Gamma_1\gamma^\mu\gamma^\nu)\otimes
				(\gamma_\nu\gamma_\mu\gamma^\alpha\gamma_\alpha\Gamma_2)$ \\
			$+ \frac{1}{24}(\Gamma_1\gamma^\mu\gamma^\nu)\otimes
				(\gamma_\nu\gamma^\alpha\gamma_\mu\gamma_\alpha\Gamma_2)$ \\
			$+ \frac{1}{24}(\Gamma_1\gamma^\mu\gamma^\nu)\otimes
				(\gamma_\nu\gamma^\alpha\gamma_\alpha\gamma_\mu\Gamma_2)$}}
						   & \multicolumn{1}{c|}{\makecell{
			$ 0 $}} \\
		\cline{1-3}
	\end{tabular}
	}
	\caption{One-loop results with an insertion of operator with structure $(\Gamma_1)~\otimes~(\Gamma_2)$ and a single
	standard model counterterm insertion. The results
	are presented immediately after loop integration and replacement of explicit counterterms with no reduction of Dirac algebra. All results are multiplied by
	$-iC_{12}(\alpha/(4\pi))^2$ and non-local divergences are neglected as they are equal to the $1/\epsilon^2$ poles. Furthermore, we set $Q_\mu=Q_e=-1$.}
	\label{tab:CTs}
\end{table}

\newpage

\section{Greek method}\label{app:greek}

In this appendix we review the greek method to reduce Dirac structures to the physical basis. Traditionally, this method is used in the literature to define EVs in higher-order calculations \cite{Tracas:1982gp,Buras:1989xd,Buras:1991jm}. As was shown in Sec.~\ref{sec:pres}, when choosing the EV basis such that the evanescent-to-physical mixing vanishes, the greek identities can simply be used to reduce Dirac structures.

\subsection{Greek identities}
In this subsection we will review how to obtain the greek identities, which can be used to either directly reduce Dirac structures to the physical basis, or to define EVs. Starting from a given Dirac structure, in order to find the corresponding greek identity, an Ansatz has to be made. In this Ansatz one assumes that the Dirac structure is proportional to its four-dimensional counterpart. The proportionality constants will contain besides the four-dimensional part also $\mathcal{O}(\epsilon^n)$ terms. To fully determine the proportionality constants the "greek trick" is used, which consists of replacing the tensor product by appropriate Dirac structures and therefore collapsing it to a single Dirac string. The constants can then be determined by reducing the Dirac strings on both sides and solving the resulting system of equations.
In order to illustrate this method, let us consider the example of the following operator:

\begin{equation}\label{eq:greekop}
  O^{3\gamma LR}=(\overline f_1 \gamma_\mu\gamma_\nu\gamma_\rho P_L f_2)(\overline f_3 \gamma^\mu\gamma^\nu\gamma^\rho P_R f_4)\,.
\end{equation}

This operator reduces in four dimensions to the standard vector operator 

\begin{equation}
O^{V,LR} = (\overline f_1 \gamma_\mu P_L f_2)(\overline f_3 \gamma^\mu P_R f_4)\,,
\end{equation}

using the Chisolm identity, which fixes the four-dimensional part of the greek identity. In four dimensions one finds:

\begin{equation}
	O^{3\gamma LR} = 4\, O^{V,LR}\,. \quad (d=4)
\end{equation}

In general $d$-dimensions however, four-dimensional Dirac relations can not be used anymore, since they are only valid up to $\mathcal{O}(\epsilon)$. Following the greek prescription we therefore make the Ansatz
\begin{equation}
	(\gamma_\mu\gamma_\nu\gamma_\rho P_L)_{ij} \otimes (\gamma^\mu\gamma^\nu\gamma^\rho P_R)_{kl}  = A\, (\gamma_\mu P_L)_{ij} \otimes (\gamma^\mu P_R)_{kl}\,,
\end{equation}
where the constant $A$ needs to be fixed. This is achieved in the Greek method by replacing the tensor product by the identity matrix, therefore collapsing the tensor product of two Dirac currents to a single one.\footnote{In principle any other Dirac structure besides the identity matrix can be used, which does not lead to a vanishing expression.} One finds:

\begin{equation}
	(\gamma_\mu\gamma_\nu\gamma_\rho P_L\gamma^\mu\gamma^\nu\gamma^\rho P_R)_{ij}  = A\, (\gamma_\mu P_L\gamma^\mu P_R)_{ij},,
\end{equation}

Both sides can now be reduced by using for example the NDR scheme. This leads to 
\begin{equation}
	(4d+(d-4)(2-d)d)( P_R)_{ij}  = A\, d(P_R)_{ij}\,,
\end{equation}
leading to the coefficient:

\begin{equation}
	A = 4(1+\epsilon)+\mathcal{O}(\epsilon^2)\,.
\end{equation}

Therefore, the greek identity for the operator structure in eq.~\eqref{eq:greekop} reads:

\begin{equation}
	(\gamma_\mu\gamma_\nu\gamma_\rho P_L)_{ij} \otimes (\gamma^\mu\gamma^\nu\gamma^\rho P_R)_{kl}  = 4(1+\epsilon)\, (\gamma_\mu P_L)_{ij} \otimes (\gamma^\mu P_R)_{kl}\,,
\end{equation}
which agrees for instance with the findings in \cite{Buras:2012fs}. 

In case the Dirac structure to reduce is proportional to two Dirac structures in  four dimensions, two proportionality constants have to be used. This is exemplified for instance by tensor structures. Considering for instance a TLL structure, the general Ansatz dictated by the four-dimensional case is given by

\begin{equation}
	(\sigma_{\mu\nu}P_L\gamma_\alpha\gamma_\beta )_{ij} \otimes (\sigma^{\mu\nu}P_L\gamma^\alpha\gamma^\beta )_{kl}  = A_T\, (P_L )_{ij} \otimes (P_L)_{kl}+ B_T\, (\sigma_{\mu\nu}P_L )_{ij} \otimes (\sigma^{\mu\nu}P_L)_{kl}\,.
\end{equation}

The constants $A_T,B_T$ are then fixed by replacing the tensor product once by the identity and once by $\gamma_{\tau_1}\gamma_{\tau_2}$. Reducing the resulting structures and solving the system of equations leads to 

\begin{equation}
	A_T = -48+80\epsilon\,,\quad B_T= 12-6\epsilon\,.
\end{equation}

Finally, for structures that do not generate chirality flips the tensor product has to be replaced by an odd number of gamma matrices for the resulting equations to be non-trivial. 

Hence, with the greek trick all Dirac structures can be mapped back to the physical basis, by fixing the scheme-dependent constants in a particular way. In our approach the factors proportional to $\epsilon$ are left general in order to be agnostic about the used $\gamma_5$-scheme.

\subsection{Greek Evanescent Operators}
In the literature the greek identities discussed in the previous subsection are typically used to introduce evanescent operators. They are typically defined by the difference of the LHS and RHS of a greek identity. For instance, for the operator in eq.~\eqref{eq:greekop}, instead of using the greek identity directly, one introduces an evanescent operator, $E^{V,LR}$, in the following way:

\begin{equation}
	E^{V,LR} := O^{3\gamma LR}-4(1+\epsilon)\, O^{V,LR}\,.
\end{equation}

This EV is then added to the basis and therefore has inserted into loop diagrams in order to renormalize it.

\bibliographystyle{JHEP}

\bibliography{refs}

\end{document}